\documentclass[12pt]{article}
\usepackage{amssymb,amsmath,amsfonts,epsfig}

\newcommand{\eq}{\begin{equation}}
\newcommand{\eqx}{\end{equation}}
\newcommand{\eqn}{\begin{eqnarray}}
\newcommand{\bi}{\begin{itemize}}
\newcommand{\eqnx}{\end{eqnarray}}
\newcommand{\ei}{\end{itemize}}
\newcommand{\ad}{a^{\dagger}}
\newcommand{\fd}{f^{\dagger}}
\newcommand{\bd}{b^{\dagger}}
\newcommand{\gd}{g^{\dagger}}
\newcommand{\Ad}{{A^{\dagger}}}
\newcommand{\Bd}{{B^{\dagger}}}

\newcommand{\ra}{\rangle}
\newcommand{\la}{\langle}

\begin{document}

\begin{titlepage}
{\hfill    CERN-PH-TH/2010-240} 

{\hfill     IFUP-TH/2010-37} 

{\hfill    TPJU - 3/2010} 

{\hfill    MPI-2010-137} 
\bigskip
\begin{flushright}
\end{flushright}
\vskip 0.8cm
\begin{center}
{\Large \bf  Dimensionally reduced  SYM$_4$ at large-$N$: \\ an intriguing Coulomb approximation} \vskip 0.8cm
{\large Daniele Dorigoni$^{a,b,c}$,
Gabriele Veneziano$^{d,e}$ and
Jacek Wosiek$^{f}$} \\[0.6cm]
{\it $^a$ Scuola Normale Superiore, Piazza dei Cavalieri 7, 56126 Pisa, Italy}\\
{\it $^b$ Department of Physics ``E.Fermi'', Pisa University, Largo B.Pontecorvo 3, Ed.C, 56127 Pisa, Italy }\\
{\it $^c$ INFN, Sezione di Pisa, Largo B.Pontecorvo 3, Ed.C, 56127 Pisa, Italy }\\
{\it $^d$ Colle\`ge de France, 11 place M. Berthelot, 75005 Paris, France} \\
{\it $^e$ Theory Division, CERN, CH-1211 Geneva 23, Switzerland} \\
{\it $^f$ Smoluchowski Institute of Physics, Jagellonian University Reymonta 4, 30-059 Cracow, Poland}
\end{center}
\begin{abstract}

We consider the  light-cone (LC) gauge and LC quantization  of the
dimensional reduction of super Yang Mills  theory from four to two
dimensions. After integrating out all unphysical degrees of freedom,
the non-local LC  Hamiltonian exhibits an explicit ${\cal N}=(2,2)$
supersymmetry. A further SUSY-preserving compactification of
LC-space on a torus of radius $R$, allows for a large-$N$ numerical
study where  the smooth large-$R$ limit of physical quantities can
be checked. As a first step, we consider a simple, yet quite rich,  ``Coulomb
approximation"  that  maintains an  ${\cal N}=(1,1)$ subgroup of the
original supersymmetry and leads to a non-trivial  generalization of
't Hooft's model with an arbitrary --but conserved--  number of
partons. We  compute numerically the eigenvalues and eigenvectors
both in momentum and in position space. Our  results, so far limited
to the sectors with 2, 3 and 4 partons, directly and quantitatively
confirm a simple physical
picture in terms of a string-like interaction with the
expected tension among pairs of nearest-neighbours  along the
single-trace characterizing the large-$N$ limit.  Although broken by
our approximation, traces of the full  ${\cal N}=(2,2)$
supersymmetry are still visible in the low-lying spectrum.

\end{abstract}
\vskip 1cm \hspace{0.7cm} October 2010
\end{titlepage}

\section{Introduction}
\label{sec1}
\setcounter{equation}{0}

Supersymmetric (SUSY) gauge theories give us a useful playground where we can test and try to understand non-perturbative features of gauge-theories.
On the other hand large-$N$ expansions of various kinds \cite{'tHooft:1973jz,Veneziano:1976wm} provide further simplifications in the
 non-perturbative dynamics to the extent that suitable combinations of  SUSY and large-$N$ often result in models that can be (almost) 
fully understood analytically and/or numerically. A well known example of such a powerful mix is, of course, the AdS/CFT correspondence \cite{Maldacena:1997re, Witten:1998qj}.

Another application of the same set of  ideas is the so-called planar equivalence \cite{Armoni:2003gp,Armoni:2003fb} between gauge 
 theories with Dirac fermions in the antisymmetric  (or  symmetric) 2-index representation of $SU(N)$ and theories with Majorana fermions 
in the adjoint. Such a correspondence (which holds at large volume for a ``common sector" in the two theories) could be physically relevant 
if $N=3$ is ``large-enough" since the former theories become, in that case, QCD-like. On the other hand, the adjoint-fermion theories become, 
in some cases, supersymmetric hence allowing to make predictions such as that  of the quark condensate in QCD from the known value of the gluino 
condensate in ${\cal N} =1$ super-Yang-Mills theory \cite{Armoni:2003yv,Armoni:2005wt}.

Furthermore, $SU(N)$ gauge theories with $N_f\geq 1$ adjoint fermions
appear to present another peculiar large-$N$ virtue, volume independence \cite{Unsal:2010qh,Kovtun:2007py},
according to which their properties at infinite volume are still captured if we consider them on a partially
compactified space  $\mathbb{R}^3\times S^1$ with periodic boundary conditions
both for the bosons and for the fermions. As the center symmetry appears to be unbroken in this case
even at small volume \cite{Unsal:2008ch}, we can send the volume of the $S^1$ to zero and yet recover,
at large-$N$, the large-volume physics (unlike the case of the well known
Eguchi-Kawai reduction \cite{Eguchi:1982nm} for fundamental fermions). Combining planar equivalence and volume independence we may dream of calculating part of the spectrum of QCD (modulo $1/N$-corrections) by a small-volume calculation in  a gauge theory with  Majorana adjoint fermions.

Having in mind these motivations, we combine in this work large-$N$ reduction and SUSY ideas by considering the well-known $\mathcal{N}=1$ supersymmetric $SU(N)$ Yang-Mills (SYM) theory compactified to two dimensions. However, since the precise connection of this model with four-dimensional SYM theory is not yet clear \cite{Unsal:2010qh}, we should consider for the time being this model for its own sake as a two-dimensional (and, to our knowledge, so far unsolved) model.

We will be using  a LC gauge and LC quantization \cite{Brodsky:1997de}, which has the advantage that all the unphysical degrees of freedom
can be integrated out at the price of introducing Dirac brackets, a common procedure for constrained systems.
The final outcome is a $2$-dimensional $\mathcal{N}=(2,2)$ supersymmetric gauge theory on the light-cone. It appears to have an exact SUSY vacuum and a well-defined, normal-ordered, positive-semidefinite  LC Hamiltonian that can be written
 as an anticommutator of supercharges.
 At large $N$  this Hamiltonian acts on color-singlet physical states  that can be written in terms of single-trace operators:
\begin{equation*}
 \mbox{Tr}\left( o_1(k_1)^\dagger...o_n(k_n)^\dagger\right)|0\rangle\,
\end{equation*}
where the different $o_i^\dagger$ will represent either bosons or fermions with different
light-cone momenta $k_i > 0$.

The full Hamiltonian is still quite involved. In this paper, after having
identified  some ``leading'' terms which present potential Coulomb-like infrared singularities, we will
define  a natural Coulomb approximation to the full Hamiltonian. We show that, in this rather drastic approximation, the model  becomes
  a partially (i.e. $\mathcal{N}=(1,1)$) supersymmetric   generalization \cite{Armoni:1995tf} of 't Hooft's  model in two dimensions \cite{'tHooft:1974hx} with non-interacting sectors characterized by the number and species of partons they contain.

  Some analytic results will be presented but most of the calculations will be numerical. This is done
 by discretizing the LC momenta \cite{Matsumura:1995kw} $k_i= n\epsilon$,  thereby reducing the problem to the diagonalization of
a large $M\times M$ hermitean matrix. We then prove that our discretization indeed converges to a continuum-limit
once we take the $\epsilon \rightarrow 0$ ($M\rightarrow \infty$)  limit.

The rest of the paper is organized as follows:
in Section 2 we sketch the derivation of the LC two-dimensional theory  starting from the SYM Lagrangian in four dimensions. We give the explicit expressions for the SUSY charges, light-cone momentum and Hamiltonian that satisfy the usual SUSY algebra.
In Section 3 we introduce our  Coulomb approximation and show how its apparent infrared divergences neatly cancel for colour-singlet states leaving behind the effects of a confining linear potential.
In Section 4 we introduce a LC compactification of the two-dimensional theory as a devise to work with discrete values of the light-cone momenta. In Section 5 we give our analytic and numerical results for the 2, 3 and 4 partons sectors of the Coulomb Hamiltonian and discuss the left-over traces of the original supersymmetry. Finally, in Section 6, we give some conclusions and a short outlook.

\section{Light-cone SYM$_4$  and its dimensional reduction}
\label{sec2}

Our starting point is $4$-dimensional $\mathcal{N}=1$  super Yang-Mills (SYM) theory as defined by the Lagrangian \cite{Terning:2006bq}:
\eq
 \label{lagSYM}
  \mathcal{L} =-\frac{1}{4} F_{\mu\nu}^a F^{a \mu\nu}+i\lambda^{a\dagger} {\bar{\sigma}}^\mu D_\mu \lambda^a\,
 =\mbox{Tr} \left[-\frac{1}{2}F_{\mu\nu}F^{\mu\nu}+{2i}\lambda^\dagger
\,D_\mu {\bar{\sigma}}^\mu\,\lambda \right]\, .
 \eqx
The gauge group is $SU(N)$ and we have set to zero a possible  $\theta$-term since it can be rotated away in the absence of a gluino mass (i.e. as long as we do not break SUSY).
We shall use the following conventions:
\begin{align*}
 F^a_{\mu\nu}=&\partial_\mu A_\nu^a-\partial_\nu A_\mu^a-g f^{abc} A_\mu^b A_\nu^c\,, ~ D_\mu \lambda^a=\partial_\mu \lambda^a-g f^{abc} A_\mu^b \lambda^c\,,\\
A_\mu =& A_\mu^aT^a_{fund}\, , ~ F_{\mu\nu}= F_{\mu\nu}^a T^a_{fund}=\partial_\mu A_\nu- \partial_\nu A_\mu+ig[A_\mu,A_\nu]\,, D_\mu \lambda=& \partial_\mu \lambda+ig [A_\mu,\lambda]\,.
\end{align*}

We now introduce LC coordinates $x^{\pm}= \frac{x^0 \pm x^3}{\sqrt{2}}$ and fix the LC  gauge $A_-=0$. As a result
we can rewrite (\ref{lagSYM}) as:
 \begin{align}
  \mathcal{L}=\mbox{Tr}&\notag\left[  (\partial_- A_+)^2+2(\partial_- A (\partial_+ A^\dagger-\bar{D} A_+)+
 \partial_- A^\dagger (\partial_+A-DA_+))+\right.\\
 &\label{lagLC} \left.(\bar{D}A-D A^\dagger-ig[A^\dagger,A])^2+{2\sqrt{2}i}\lambda^\dagger
  \left(
 \begin{matrix}
 \partial_- & {-}\bar{D}\\
 {-}D &  D_+
  \end{matrix}
  \right) \lambda\right]\,,
 \end{align}
where we have introduced $A = (A_1+i A_2)/\sqrt{2}  ,\, A^\dagger = (A_1-iA_2)/\sqrt{2} $,  $D_{1+i2}\equiv \sqrt{2} D=\sqrt{2}(\partial+ig[A,\times])$ and its hermitean conjugate
$D_{1-i2}\equiv \sqrt{2} \bar{D}=\sqrt{2}(\bar{\partial}+ig[A^\dagger,\times])$.

The fields $A_+$ and $\lambda_1$ become non-dynamical (in the sense that their equations of motion
do not involve derivatives with respect to the lightcone time $x^+$) and can be integrated out. Furthermore, we can consider the reduction of the theory to $D=2$ by discarding all dependencies on $x^\perp$ and hence by setting to zero $\partial$ and $\bar{\partial}$. As a result  the Lagrangian under consideration becomes:
\begin{align}
 \mathcal{L}^r_{io}=&\notag\mbox{Tr} \left[J_r^+ \frac{1}{\partial_-^2} J_r^+ +2(\partial_- A\partial_+ A^\dagger + \partial_+ A\partial_-A^\dagger)-g^2 [A,A^\dagger]^2+\right.\\
&\label{Liored}+\left.{ 2\sqrt{2}i}(\lambda_2^\ast \partial_+ \lambda_2-g^2[A,\lambda_2^\ast]\frac{1}{\partial_-} [A^\dagger,\lambda_2]) \right]\, ,
\end{align}
where the subscript $io$ means that we have integrated out the non-dynamical fields and the superscript $r$ tells us that this is  now a theory in $2$-dimensions $x^{\pm}$. The (reduced) current $ J_r^+$ turns out to be:
\begin{equation}
 J_r^+=ig[A^\dagger,\partial_- A]+ig[A,\partial_- A^\dagger]+{ \sqrt{2}}\,g\{\lambda_2^*,\lambda_2\}\,.\label{CurrentReduced}
\end{equation}
To obtain the Hamiltonian from (\ref{Liored}) we have simply to perform a Legendre transform:
\begin{align}
 H^r_{io}=&\label{HamiltIO}\mbox{Tr} \left[-J_r^+ \frac{1}{\partial_-^2} J_r^+ +g^2\,[A,A^\dagger]^2 + 2\sqrt{2}i\,g^2[A,\lambda_2^\ast]\frac{1}{\partial_-} [A^\dagger,\lambda_2] \right]\,,
\end{align}

We have to keep in mind, however,  that  the right commutation relations are those obtained
 from Dirac --rather than Poisson-- brackets: only then the constraints implicitly used will be preserved by the LC-time evolution generated by
 $H^r_{io}$. This leads to the following quantization of the dynamical fields $A$ and $\lambda_2$
 in terms of creation/annihilation operators with a Fock vacuum $|0\rangle$\footnote{We have assumed a vanishing VEV for the scalar field although, even in $D=2$, we expect the classical moduli space to be preserved since tunneling should be suppressed  in the large-$N$ limit. If so the spectrum will depend on the particular point chosen in moduli space. One of us (GV) is grateful to A. Armoni and A. Schwimmer for a discussion concerning this point.}:
\begin{align}
 A^a(0,x^-)=&\notag \int_0^\infty \frac{dk^+}{\sqrt{2\pi}\,\sqrt{2k^+}} \left[
a^a(k^+)e^{-ik^+ x^-}+ b^{\dagger a}(k^+)e^{+ik^+ x^-}\right]\,,\\
\lambda^a_2(0,x^-)=&\notag\int_0^\infty \frac{dk^+}{{ 2^{1/4}\sqrt{2\pi}}} \left[
f^a(k^+)e^{-ik^+ x^-}+ g^{\dagger a}(k^+)e^{+ik^+ x^-}\right]\,,\\
[a^a(k^+),a^{\dagger b}(k'^+)]=&\label{modes}[b^a,b^{\dagger b}]=\{f^a,f^{\dagger b}\}=\{g^a,g^{ \dagger b}\}=\delta(k^+-k'^+)\delta^{ab}\,,
\end{align}
and of course all other (anti-)commutators vanish. Note that in order to obtain precisely the Dirac commutation relations we have to take our LC momenta $k^+$ to be positive.

The theory thus obtained can be seen as  the  $\mathcal{N}=(2,2)$ two-dimensional supersymmetric theory that follows from the dimensional reduction
of $\mathcal{N}=1$ super Yang-Mills on $\mathbb{R}^2 \times T^2$.
Note, however, that due to the non-susy invariance of the gauge fixing the transformations
generated by the supercharges through Dirac brackets will not be the usual one but will
have to be supplemented by
an additional gauge transformation needed to restore the gauge constraint $A_-=0$. This is already true before compactification (where one gets  $\mathcal{N}=1$ SUSY in $D=4$) and keeps working  after compactifying on $\mathbb{R}^2 \times T^2$ i.e. if we keep
only the zero-modes (hence setting $\partial/\partial x^\perp=0$).
A straightforward calculation leads to the following form for the supersymmetric charges, momentum and Hamiltonian operators:

\begin{align}
 Q_2  &=\notag \int dx^-\,  2\sqrt{2} \partial_- A^a \lambda_2^{\ast a}\\
&= 2^{3/4} \int_0^\infty dk_1 dk_2 \,\sqrt{k_1} \delta(.) \,\left(
b_{k_1}^{a\dagger} g_{k_2}^a -a_{k_1}^a f_{k_2}^{a\dagger}
\right)\,,\\
P_-&\notag= \frac{1}{2\sqrt{2}} \{Q_2,{\bar{Q}}_{\dot{2}}\}=\int dx^- \, (2 \partial_- A^{\dagger a} \partial_- A^a+i\sqrt{2} \lambda_2^{\ast a} \partial_- \lambda_2^a)\\
& =\int_0^\infty dk \,k \,\left[a^{\dagger a}_{k} a^a_{k} + b^{\dagger a}_{k} b^a_{k} + f^{\dagger a}_{k} f^a_{k} +g^{\dagger a}_{k} g^a_{k}\right]\,, \\
 Q_1 &=\notag \int d x^- (-2\sqrt{2} g f^{abc} A^{\dagger a}
 \partial_- A^b+2ig f^{abc}\lambda_2^a \lambda_2^{\ast b}) \frac{1}{\partial_-}\lambda_2^{\ast c}\\
&=\notag \frac{g}{2^{1/4}\sqrt{\pi} } \int_0^\infty dk_1 dk_2 dk_3\, \delta(.)f^{abc}
\left[ \frac{\sqrt{k_2}}{k_3 \sqrt{k_1}} \left( a_{k_1}^{a\dagger}a_{k_2}^b g_{k_3}^c -a_{k_1}^{a\dagger} a_{k_2}^b f_{k_3}^{c\dagger}\right. \right.\\
&\notag \left. + b_{k_1}^a b_{k_2}^{b \dagger} f_{k_3}^{c\dagger}-b_{k_1}^a b_{k_2}^{b\dagger} g_{k_3}^c\right) +\frac{\sqrt{k_1}}{k_3\sqrt{k_2}}  \left( b_{k_1}^{a\dagger} a_{k_2}^{b\dagger} g_{k_3}^c +a_{k_1}^a b_{k_2}^b f_{k_3}^{c\dagger}\right) \\
&\label{Q1}+ \frac{(k_3+k_2)}{2 k_3 k_2} \left( g_{k_1}^{a\dagger} g_{k_2}^b g_{k_3}^c-f_{k_1}^a f_{k_2}^{b\dagger} f_{k_3}^{c\dagger}\right) \left. + \frac{(k_2-k_3)}{k_3 k_2} \left(f_{k_1}^a f_{k_2}^{b\dagger} g_{k_3}^c+g_{k_1}^{a\dagger} f_{k_2}^{b\dagger} g_{k_3}^c\right) \right]
 \\
 P_+&= \frac{1}{2\sqrt{2}} \{Q_1, \bar{Q}_1\}= \int dx^-\, H^r_{io}\,,
\end{align}
where $\delta(.)$ is a shorthand notation for $\delta(\pm k_1 \pm ...)$ where the sign plus is for creation operator while the minus sign otherwise.
We did not report here the explicit  momentum space version for $P_+$ since, as the reader may guess from the rather involved expression for $Q_1$, its expression is quite long and not particularly illuminating. A lengthy calculation shows that the above charges
 satisfy the $\mathcal{N}=(2,2)$ SUSY algebra:
\begin{align}
 \{Q_1,\bar{Q}_1\} =& \notag 2\sqrt{2} P_+=2\sqrt{2} H_{io}^r \,,\\
\{Q_2,\bar{Q}_2\} =&\label{reducedSusy} 2\sqrt{2} P_- \,,
\end{align}
with all other anticommutators vanishing identically.

As we will see in the following sections one can compactify also the $x^-$ direction and replace all the integrals $\int_0^\infty dk$
with sums over positive integers  $\sum_{n=1}^\infty$, the key point is precisely that all LC momenta have to be greater than zero, so we can
actually rewrite $\int_0^\infty dk= \int_{-\infty}^{+\infty}dk\, \theta(k) $.
We note, incidentally that in \cite{Harada:2004ck} it was claimed that the discretized version
of the supercharges does not satisfy the susy algebra.
We claim instead that everything works fine provided an appropriate care is used in performing
the discretization and in defining the (anti)commutators;  particular attention is needed when, by momentum conservation, an intermediate parton gets a vanishing momentum: by adopting a careful prescription for the ensuing $\theta(0)$ we can fully maintain supersymmetry.

\subsection{General properties of the reduced theory}

Let us discuss some exact features of the complete SUSY charges and Hamiltonian.
The Hamiltonian conserves parton number parity  $(-1)^p$ (actually changes $p$ by $0, \pm 2$)
and therefore splits into two sectors with even and odd number of partons.
On the other hand the charges $Q_2, Q_2^{\dagger}$ preserve $p$ while $Q_1, Q_1^{\dagger}$
change $p$ by $\pm 1$.

 Besides $P_-$ there is another conserved quantity, $J_z$, which is
 what remains of the original helicity of the $4$-D theory.
 The four partons $a,b,f,g$ have $J_z = +1, -1, +1/2, -1/2$ respectively and the hamiltonian is block-diagonal with subspaces of fixed total $J_z$ while  $Q_1, Q_2^{\dagger}$ change $J_z$ by $+1$ and $Q_2, Q_1^{\dagger}$ by $-1$.

 Consider for instance a supermultiplet containing a $J_z=0$ boson $|x\rangle$ of non-vanishing mass. Since its $p_+$ and $p_-$ are both non-vanishing this state can be at most annihilated by one out of  $Q_1, Q_1^{\dagger}$ and similarly by one out of  $Q_2, Q_2^{\dagger}$. Acting on $|x\rangle$ in all possible ways we see that we generate a supermultiplet  containing two bosons with opposite parton number parity and two fermions with opposite parton parity.
 However CPT invariance requires the existence of a similar multiplet of antiparticles and we end up with 4 bosons and 4 fermions as the minimal size of a massive supermultiplet\footnote{This is only apparently in contrast with the counting of states in $D=4$ but it is not since, in $D=2$ there is a further doubling of states due to to the distinction between left- and right-movers \cite{West:1990tg}. }.

 As an example, consider the anomaly (or Konishi) supermultiplet $S \equiv \epsilon^{\alpha \beta} W_{\alpha}W_{\beta}$. Its lowest component is the gaugino bilinear $\epsilon^{\alpha \beta} \lambda_{\alpha}\lambda_{\beta}$ which is just proportional to $\lambda_1 \lambda_2$. Since in our setup  $\lambda_1$ is expressed in terms of
 $\lambda_2$ and $\bar{A}$ we see that, at first order,  $S|0\rangle$ gives a state of the type $|gga\rangle$. Such a state is clearly annihilated by $\bar{Q}_2$ (but not by
 $Q_2$) and one can consistently assume that it is annihilated by $\bar{Q}_1$ (and not by $Q_1$), which mimics exactly the situation in $D=4$.
 Acting on $|gga\rangle$ with $Q_1$, $Q_2$ and $[Q_1,Q_2]$ (the anticommutator being zero), we generate two fermionic states (whose lowest $p$ components look like
  $|ga\rangle$ and a combination of $|ggf\rangle$ and $|gab\rangle$) and one more scalar (a combination of $|ab\rangle$ and $|fg\rangle$ with additional  $p>2$ components).
  This gives a total of 2 bosons and 2 fermions to which we have to add a similar set of states starting from $\bar{S}|0\rangle \sim |ffb\rangle$.

 As we shall see, our approximation to the full Hamiltonian breaks SUSY to an ${\cal N}=(1,1)$ subgroup of the full  ${\cal N}=(2,2)$ and therefore to the breaking of some degeneracies. Nonetheless, an approximate full degeneracy will be seen to remain even in our very simple system with Coulomb interactions only.

\section{Cancellation of leading infrared divergences and  a Coulomb approximation}
\label{sec3}

In our LC quantization  the supercharge $Q_2$ and the momentum $P_-$ are like in a free theory and thus trivial. By contrast, the supercharge $Q_1$ and the ``Hamiltonian" $P_+$ are highly not trivial:  non-linear and even non-local. Let us discuss some of their properties before making any approximation.
From (\ref{Q1}) we see that every term appearing in $Q_1$ is trilinear in creation/annihilation operators. Furthermore, since the LC momenta are all positive,  momentum conservation implies the absence of  pure creation or pure annihilation terms. Thus $Q_1$ connects states with opposite parton-number parity.  The LC Hamiltonian,
$\{Q_1,\bar{Q}_1\}$,  will have either quadratic terms (with one creation and one annihilation operator), or quartic terms. Since no pure destruction or pure creation operators appear in $H_{io}^r$  the Fock vacuum is annihilated both by the SUSY charges and by the Hamiltonian and is an exact zero-energy ground state. The quartic terms induce either $2 \rightarrow  2$ or $1 \rightarrow  3, 3 \rightarrow  1$ transitions and thus  conserve  $(-1)^p$.

The full Hamiltonian exhibits, at least superficially, both linear and logarithmic
infrared (IR) divergences when some momenta (or combinations thereof) go to zero.
The logarithmic divergences, that resemble those of the $4D$ theory,
presumably need a Block-Nordsieck treatment that we plan to
implement (analytically and/or numerically) in a forthcoming paper.
The linear, Coulomb-like divergences are instead neatly cancelled for colour-singlet states as we shall now argue. Nevertheless, the finite effects they leave behind are expected to dominate the Hamiltonian at large distances.

In order to illustrate this feature we introduce a rather crude
``Coulomb approximation",  in which we  keep only the linearly-divergent terms in the Hamiltonian in their minimal form of $1/q^2$ poles. Such IR singularities only appear in the quadratic terms (which are necessarily diagonal) and in  the elastic  $2 \rightarrow  2$ scattering terms. In the above-mentioned approximation the former take the simplified  form:

\begin{equation}
 H_C^{quad} = \frac{\lambda}{\pi}\sum _A \mbox{Tr}
 \int_0^\infty dk\int_0^k \frac{dq}{q^2} A_k^\dagger A_k~~;~~ \lambda \equiv  g^2 N_c~~,
\label{Hquadratic}
\end{equation}
where $A$ stands for any one of the four parton species. Similarly the elastic quartic terms  can be simplified as:

\begin{align}
 H_C^{elastic} =  - \sum_{A,B}
 \frac{g^2}{2\pi} \int_0^\infty dp_1 dp_2 &\notag \left[
\int_0^{p_1} \frac{dq}{q^2} \mbox{Tr} (A_{p_1}^\dagger B_{p_2}^\dagger B_{p_2+q} A_{p_1-q}) +\right.\\
&+\left. \int_0^{p_2}\frac{dq}{q^2} \mbox{Tr} (A_{p_1}^\dagger
B_{p_2}^\dagger B_{p_2-q} A_{p_1+q}) \right]\,, \label{Hquartic}
\end{align}
where $A, B$ stand for any one of the four parton species. Note that
these elastic terms neither change the number of partons nor their
species. For this reason we can restrict ourselves to subspaces of
the entire Hilbert space of states with a given set (i.e. number and
species) of partons and fixed total momentum $P_- (\equiv P$ below).
Furthermore, as discussed in Section 1, the 't Hooft limit, $N
\rightarrow \infty$ with $\lambda$ fixed, selects
single trace states as the only relevant ones (transitions with
multi-trace states being $1/N$-suppressed).

In conclusion, we can diagonalize $H_C$   by splitting the total
Hilbert space generated by a generic linear superposition of
single-trace  states  into subspaces of definite total momentum $P$
and definite  parton number  $p$:
\begin{equation}
 \mathcal{H}_P^p = \left\lbrace |s\rangle = \mbox{Tr}\left(o_1^\dagger(k_1) o_2^\dagger(k_2)...o^\dagger_{p-1}(k_{p-1})o_p^\dagger (P-\sum_{i=1}^{p-1} k_i)\right)
|0\rangle \right\rbrace\,,\label{HilbertSubspace}
\end{equation}
where $o_i^\dagger$ can be any of our creation operators and all the
momenta $k_i$ satisfy $0<k_i< P$ for $i=1,...,p$ and we  define
$k_p=P-\sum_{i=1}^{p-1} k_i$. Starting with section 5 we will study
in detail,  both analytically and numerically,  the eigenvalues and
eigenvectors of $H_C$ restricted to
$\mathcal{H}_P^2,\,\mathcal{H}_P^3$ and $\mathcal{H}_P^4$. For the
numerical approach  it will be convenient  to discretize the LC
momenta by compactifying the coordinate $x^-$. This will be
discussed in the next section.

Before turning to actual calculations let us give a general argument for the cancellation
of the Coulomb divergences for a general state of the form (\ref{HilbertSubspace}).
Let us take an arbitrary pair of neighbour partons in (\ref{HilbertSubspace}) and consider four distinct contributions to their mutual and self-interaction.
The self interactions come from the quadratic terms (\ref{Hquadratic}) for each one of the neighbour partons. However, in order to keep the book-keeping right, we attribute half of the quadratic term acting on each parton to its interaction with the left neighbour and half to the one with its right neighbour.
The book-keeping is easier for the quartic Hamiltonian where one just keeps the terms corresponding to the exchange of quanta between the two selected partons (there are two such contributions, in general, depending on which of the two partons  gains energy in the process).

When these four contributions are added one finds, for each pair of neighbours, the following result:
\begin{equation}
H_C^{A-B} \varphi(\dots p_A, p_B,\dots) =  \frac{\lambda}{2 \pi} \int_{-p_{B} +\epsilon}^{p_A - \epsilon} dq  \frac{ \varphi(\dots p_A,p_{B}, \dots ) - \varphi(  \dots p_A-q,p_{B}+q, \dots )}{q^2} \, ,
\end{equation}
where $A-B$ stand for a pair of neighbours and $\varphi(p_1,...,p_N)$ is the wavefunction in momentum space of the particular chosen state. The total Hamiltonian is given then by a sum over $A$ and $B$.
We see  that the numerator on the r.h.s.  vanishes at $q=0$. If we take a smooth linear superposition of parton-momentum eigenstates, the numerator will vanish quadratically in $q$ thus completely removing the singularity.

As a physical example of such a smooth superposition let us consider two partons displaced by some
$\Delta x  \equiv  \Delta x^-$ in LC space.
The momentum wavefunctions will combine to give an $e^{i q \Delta x}$ for the second term in the numerator
so that, for each pair:

\begin{equation}
H_C^{A-B} \simeq  \frac{\lambda}{2 \pi} \int_{-p_B}^{p_A} dq \frac{(1- cos(q\Delta x) )}{q^2}
\simeq \frac{\lambda}{ \pi} \int_{-\infty}^{\infty} dq \frac{sin^2(q\Delta x/2)}{q^2} =
 \frac{\lambda}{ 2} |\Delta x| \, ,
\end{equation}
at large separations. We have just discovered that the Coulomb Hamiltonian, acting on a single trace state,
produces an energy proportional to the sum of the $|\Delta x^-_{ij}|$ distances for each pair of
neighbouring partons. The constant of proportionality turns out to be just $\frac{\lambda}{2}$ i.e.
nothing but the $2D$ string tension for sources in the fundamental representation
\cite{Kazakov:1980zi, Kazakov:1980zj}\footnote{This can be generalized to finite $N$ with $C_F \equiv (N^2-1)/2N$ replacing $N/2$ (private communication by G. C. Rossi).}. This is as it should be, since each parton behaves, vis-a-vis of its neighbours, like a parton in the fundamental representation. Its belonging to the adjoint representation reveals itself in the fact that it has {\it two} neighbours. For a two parton system this just produces the well known factor two difference between adjoint and fundamental tension in the large-$N$ limit. A numerical verification of this analytic argument will be given in Sect. 6.

\section{Light-cone compactification}
\label{sec4}
 We now further compactify, for computational convenience,
the $x^-$ direction on a circle of radius $R$ (with periodic boundary conditions).
As a consequence, the $p_-$ momenta are quantized\footnote{In this section we shall
keep Planck's constant explicit  in order to illustrate the emergence of a quantum-mechanical string.}:
\eq
\label{mq}
p_- = n\frac{\hbar}{R} \equiv n \epsilon\, , \, n= 0, 1, \dots
\eqx
The total conserved momentum is taken to be
\eq
P_{-}\equiv P = K \frac{\hbar}{R} \equiv K \epsilon
\eqx
and we shall be interested in the decompactification limit
$R \rightarrow \infty\; (  \epsilon \rightarrow 0 ) $. In order to keep $P$ fixed this also means $K \rightarrow \infty$.
Since in LC quantization all  momenta are positive,
 $P$ effectively plays the role of an ultraviolet cut off $\Lambda_{UV}$.
 In other words, in eq. (\ref{mq}), $n \le K$.  If we also take
 out the zero mode ($n=0$ in (\ref{mq})), $\epsilon \equiv \hbar/R$ plays instead the role of
 an IR cutoff and
$K = \frac{\Lambda_{UV}}{\Lambda_{IR}}$.
The picture is similar to that of a (2-dimensional) lattice gauge theory in Hamiltonian formalism
(continuous time and discretized space)
with a lattice spacing given by $a = 2\pi \hbar / P = L/K$ and a total of $ K$ lattice points
 to cover a large circle of circumference $L = 2 \pi R = K a$.

According to the LC quantization philosophy, $P_+$ plays the role of the Hamiltonian and the Lorentz-invariant eigenvalue equation reads:
\eq
M^2 |s\rangle=2 P_+  P_-|s\rangle  = 2 P P_+ |s\rangle= 2 K \epsilon P_+|s\rangle
\eqx
The eigenvalues (and eigenvectors) of this operator should have a finite limit
as $K  \rightarrow \infty$.
We have found it convenient to work in units in which $\epsilon =1$ (integer parton momenta).
In that case the operator $K P_+$ should approach
a finite limit as $K \rightarrow \infty$. More physically we could have chosen units in which
$a=1$ and momenta are quantized in units of $2 \pi/K$.
One can easily check that the limit  $K \rightarrow \infty$  provides in both cases the same spectrum
for $M^2$.

\begin{figure}[h!]
\epsfig{width=16cm,file=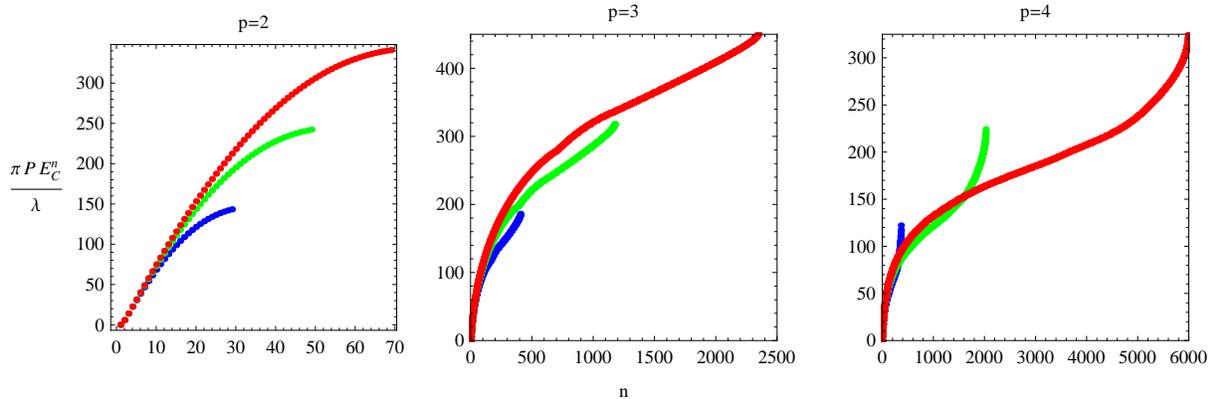}
\caption{Cutoff dependence of the eigenvalues of $H_C$ in the three lowest multiplicity sectors. The size of the $H_C$ matrix
grows with the resolution parameter $K$ and can be read from the range covered on the horizontal axis,
$K=30,50,70$ for $p=2,3$, and $K=15,25,35$ for $p=4$. }
\label{p234partonEnergies}
\end{figure}

We expect to find finite-size effects at finite $K$. Because of
periodicity the maximal $x^-$ distance is $ K a / 2$ and therefore
only states whose wavefunction is concentrated in regions much
smaller than $ K a/2$ are expected to have already reached an
asymptotic limit at some given $K$. We will see later that this is
exactly what happens numerically, but let us anticipate the presentation of
some of the results we will find for $p=2$. Once we have
discretized the momenta in units of $\epsilon \equiv 1$, the $M^2$
operator  becomes an $( K - 1)\times ( K - 1) $ matrix whose low
eigenvalues  converge fast to some finite values while increasing
$K$. This can be seen in Fig. \ref{p234partonEnergies} (to be
discussed in detail in Section  \ref{Results}) which illustrates
what happens for
 $K=30,50$ and $70$. This is equivalent to discretizing states with a fixed total momentum
 $P=1$ by subsequently taking
$\epsilon=1/30,1/50$ and $1/70$, We see a neat convergence of the lowest eigenvalues
towards a smooth (and almost linear) spectrum.

Referring still to the two-parton sector, we shall also find that the eigenvectors exhibit a sharp localization in position  (i.e. $x^-$) space, or, more precisely, in the distance  $\Delta x^-$ between the two partons. The average distance appears to be quantized:
\eq
\label{ad}
\langle \Delta x^- \rangle_n = c_n \frac{L}{K} = 2 \pi c_n \frac{\hbar}{P_-}~~~,~~     n = 1, 2, \dots   K-1 \, ,
\eqx
where the $c_n$ are a sequence of numbers going from a number of $O(1)$  to a number of $O( K/2)$
in steps of $O(1)$. The ordering  coincides with that of the energy eigenvalues.
In other words, the average distance is of the order of $a = 2\pi\hbar/P_-$ for the lightest states
and grows up to the maximal allowed physical distance for the heaviest eigenstates.
As we increase $K$, the low-lying states stabilize while the heavy ones change and stabilize only at higher values of $K$ (i.e. when $\Delta x^-$ is well within the compact circle). The spread in $\Delta x^-$ is $O(a)$ for all the states (this is the above-mentioned position-space localization).

The low-lying eigenvalues of $P_+$ behave like $P_{-}^{-1}$ for large $P_{-}$ and, indeed,
one finds (c.f. Fig.\ref{E2(l)}) an approximate linear relation between energy and average distance:
\eq
\label{edr}
P_+^{(n)} \simeq \lambda  \langle \Delta x^- \rangle_n
\eqx
where $\lambda $ is 't Hooft's coupling normalized in such a way that it corresponds to a classical tension. This means that an appealing string picture emerges whereby energies are proportional to the string tension $\lambda$. This result, however,  is Lorentz-frame dependent. In order to find a Lorentz invariant result we compute the mass eigenvalues:
\eq
\label{mdr}
M_n^2 = 2P_- P_+^{(n)} \simeq 2 \lambda P_-  \langle \Delta x^- \rangle_n  = 4 \pi  \hbar \lambda c_n
=  \lambda M_n  \langle \Delta x^{cm}\rangle_n
\eqx
where we have used first (\ref{ad}) and, for the last step, an $n$-dependent Lorentz transformation
with boost $\sqrt{2}P_-/M_n$ in order to go to the $n$th state  rest frame. Thus we finally  obtain:
\eq
\label{m2dr}
\langle \Delta x^{cm}\rangle_n    =  \sqrt{4 \pi c_n} l_s ~~, ~~
M_n \simeq  \lambda \langle \Delta x^{cm}\rangle_n    =   \sqrt{4 \pi c_n} M_s
\eqx
in terms of the string length and mass scales: $l_s = \sqrt{\hbar/\lambda}, ~ M_s \equiv \sqrt{\hbar \lambda}$. This is in perfect  agreement with expectations\cite{Kazakov:1980zi, Kazakov:1980zj} once we realize that, for $p=2$, the sum over neighbour-parton  distances is $2 \Delta x$.

\section{Bases and matrix elements at large $N_c$}

Below we quote a few explicit expressions for matrix elements of the discretized (and rescaled) Hamiltonian $h_C$ defined by:
\eqn
H_C = \frac{\lambda}{\pi}\frac{K}{P}h_C.
\eqnx
The mode expansion of $h_C$ in terms of the discretized creation and annihilation operators
\eqn
A_m = \frac{1}{\sqrt{R}} A_{m\epsilon}
\eqnx
is the same as in (\ref{Hquartic},\ref{Hquadratic}) with all momenta $p\rightarrow m$
and integrals $\int dp \rightarrow \sum_m$.

There are four kinds (species) of partons in our model: two bosons and two fermions
with the corresponding annihilation operators denoted by $a,b,f,g$. The detailed construction
of the planar bases in each multiplicity sector differs slightly depending on the  particular choice of parton
species involved and the same applies to the matrix elements of $h_C$ (\ref{Hquadratic},\ref{Hquartic}).

\subsection{Two different partons}

At given resolution $K$ the states belonging to the discretised
version of $\mathcal{H}^{2}_K$ are labeled by one integer:
\eqn
|n \ra = Tr[\Ad_n \Bd_{K-n}]|0\ra,\;\;\; n=1,...,K-1\,,
\eqnx
where  $A,B=a,b,f,g$. The large N rules, developed e.g. in \cite{Veneziano:2005qs}, give
\begin{align}
\la n | h_C | n' \ra =&\notag  \delta_{n,n'} \left(  C(n) +  C(K-n)\right )
- (1-\delta_{n,n'})\frac{1}{2(n-n')^2}
\\
&\label{p2H} -(1-\delta_{K-n,K-n'})\frac{1}{2(n-n')^2}\,,  \\
 C(n)&\notag=\sum_{q=1}^{n-1} \frac{1}{q^2}\,.
\end{align}
\subsection{Two identical partons}
In this case, due to the cyclic symmetry $n \rightarrow K - n$, only half of the states from above,
is  linearly independent. The matrix elements can be neatly expressed by (anti)symmetrized
ones from the previous case
\begin{equation}
 \la n | h_C | n'\ra_{id}= \la n | h_C | n' \ra_{dif} \pm \la K-n | h_C |n'\ra_{dif}\,,\,\,\,\,\,
 n=1,...\frac{K-1}{2},\,\,\,K\, odd,
\end{equation}
where the suffix $id$ stands for identical partons and the suffix $dif$ for different partons while the $+$ sign is for identical bosons and the $-$ for fermions.
The union of the two (bosonic and fermionic) spectra for identical partons will reconstruct
the complete spectrum obtained from eq.(\ref{p2H}).

\subsection{Three different partons}

We consider now states composed by three different partons (taken as {\em abf }). These  are labeled by two integers:
\eqn
|n,m \ra = Tr[\ad_n \bd_m \fd_{K-n-m}]|0\ra,\;\;\; 1\le n \le K-2,\,\,\, 1 \le m \le K-n-1
\eqnx
and the matrix elements of Eq. (\ref{Hquadratic},\ref{Hquartic}) read:
\begin{align}
\la m,n |h_C |m',n'\ra=&\notag \delta_{m,m'}\delta_{n,n'}(C(m)+C(n)+C(K-m-n)) \\
                      &\notag- \delta_{m,m'}(1-\delta_{n,n'})/2(n-n')^2 \\
                      &\notag- \delta_{n,n'}(1-\delta_{m,m'})/2(m-m')^2 \\
                      &\label{p3H}- \delta_{m+n,m'+n'}(1-\delta_{m,m'})/2(m-m')^2\, .
\end{align}
In the case of states with just two parton species (e.g. $aaf$) $h_C$ has  the same matrix elements.
The Hilbert space is however twice as small since cyclic and anticyclic permutations correspond
to the same state. On the other hand the bigger Hilbert space for three different species splits into
two separate sectors, since states corresponding to above cyclic and anticyclic permutation do not interact
in the planar limit.

\subsection{Three identical partons}

With a trick similar to the two parton case one can exploit the cyclic symmetry and express matrix elements in terms of  those
for three different partons

\begin{align}
 \la n,m| h_C |n',m'\ra_{id}=&\notag\la n,m| h_C |n',m'\ra_{dif}+\la m,K-n-m| h_C |n',m'\ra_{dif}+\\
&\label{p3idH}+\la K-n-m,n| h_C |n',m'\ra_{dif}\,.
\end{align}
This time there is no difference between bosonic and fermionic
sectors since $\mathbb{Z}_3$ shifts involve an even number of fermionic
transpositions.
\subsection{Four different partons, the {\em abfg} sector}.

The planar basis is now parametrized by three integers

\begin{align*}
&|n,m,o \ra =Tr[\ad_n \bd_m \fd_o \gd_{K-n-m-o}]|0\ra\,,\\
& 1\le n \le K-3\,,\,\,\,\, 1 \le m \le K-n-2\,,\,\,\,\,1 \le o \le K-m-n-1\, ,
\end{align*}
and matrix elements of Eq. (\ref{Hquadratic},\ref{Hquartic}) read
\begin{align}
\la m,n,o |h_C |m',n',o'\ra=&\notag
\delta_{m,m'}\delta_{n,n'}\delta_{o,o'}(C(m)+C(n)+C(o)+C(K-m-n-o)) \\
&-\notag \delta_{m,m'}\delta_{n,n'}(1-\delta_{o,o'})/2(o-o')^2 \\
&-\notag \delta_{n,n'}\delta_{o,o'}(1-\delta_{m,m'})/2(m-m')^2 \\
&-\notag \delta_{o,o'}\delta_{m+n+o,m'+n'+o'}(1-\delta_{n,n'})/2(n-n')^2 \\
&\label{p4H} - \delta_{m+n+o,m'+n'+o'}\delta_{m,m'}(1-\delta_{o,o'})/2(o-o')^2\,.
\end{align}

Generalization to higher parton multiplicities is pretty
straightforward.

\section{Results}
\label{Results} We turn now to discuss quantitative solutions
obtained, mainly numerically, with the aid of Mathematica. The crude
Coulomb approximation introduced above is not only the lowest
approximation for the full set of the complete coupled LC
eigenequations but, at the same time, it defines a natural
generalization of 't Hooft equation to many-body sectors. Rather than
concentrate on separate parton multiplicities, $p$, we shall focus
on a few physical issues and compare results for different
multiplicities. Until now only the first three nontrivial sectors,
i.e. $p=2,3,4$ have been looked upon. Obviously, further increase of
$p$ is technically more challenging, however it is feasible, if such
a need arises, by employing more dedicated methods and algorithms.

\subsection{$K$ dependence vs. $K=\infty$ limit}

Figure \ref{p234partonEnergies}  summarizes the cutoff dependence of spectra of $H_C$
for different multiplicities. As expected, lower states converge faster, however one should
remember that at the high end of the spectrum new states appear for each $K$.
Since $K$ controls the length $L=2\pi K/P$ of our torus, the highest state, for example, will never converge because it is
a new state with higher and higher energy as we increase $K$.

The eigenenergies of $P H_C$ (that is $M_n^2/2$) are
 displayed as a function of a single index $n$ which labels consecutively ordered eigenvalues.
Only for the two parton case $n$ is directly related with a single physical observable (see below).
For more partons, $n$ is effectively composed of more quantum numbers related to other, yet unknown,
quantities conserved in a particular many body sector. As a consequence, the $p=2$ dependence
is nicely linear, at large $K$, while the limiting curves for higher $p$ are more complicated. They
reflect the above degeneracy with more and more states below a fixed energy as we increase $p$.
In fact one can read the spectral density of states $dN/dE$ directly from the figure. It grows
with $E$ at  intermediate energies, the growth becoming more and more rapid as we increase $p$.
However at the highest end of the spectrum the density saturates revealing some sort of blocking
related to periodicity.

\begin{figure}[t]
\begin{center}
\epsfig{width=18cm,file=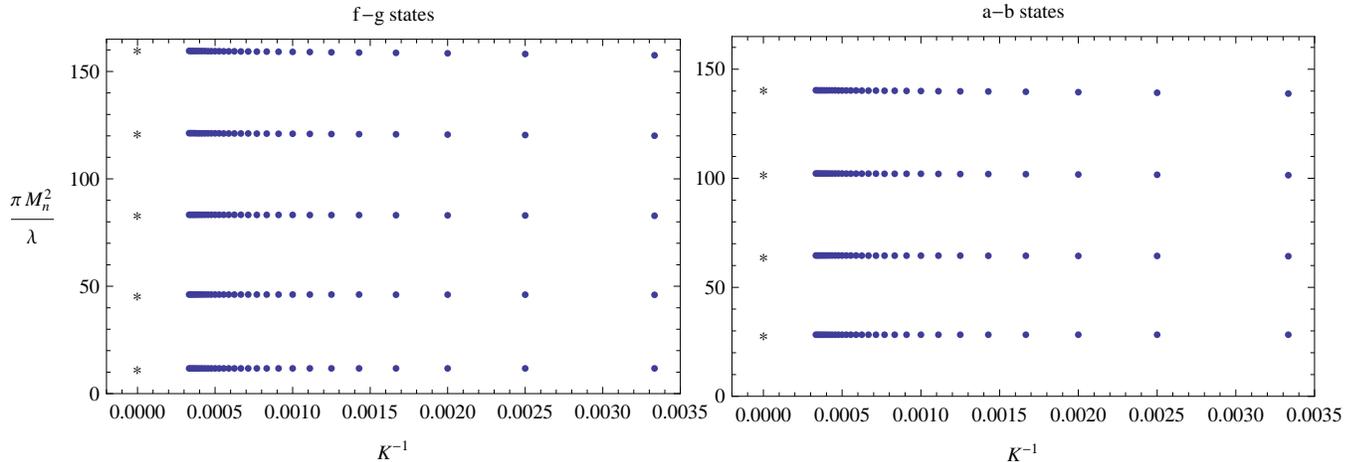}
\end{center}
\vskip-10mm \caption{Convergence of the lowest energy levels with the
cutoff $( 100 \le K \le 3000)$ and comparison between the even
(identical boson-boson) and odd (identical fermion-fermion) solutions. Stars represent direct
solutions of the 't Hooft equations for continuous momenta
($K=\infty$).}
  \label{ab-fg}
\end{figure}
Let us now look for the cutoff dependence of the lowest levels in more detail.
Figure \ref{ab-fg} shows the first few eigenenergies in the boson-boson and fermion-fermion,
$p=2$, sectors as a function of $1/K$. Evidently dependence on K is very weak and
the extrapolation to the continuum momentum limit is straightforward. In fact, in this limit the LC
eigenequations are nothing but 't Hooft equations with adjoint charges. Hence the limiting values
can be obtained by solving directly these equations. They are also displayed in Figure \ref{ab-fg},
providing a rather satisfactory cross check of the whole approach.

Finally we comment on the difference between bound states made of fermionic and/or bosonic identical partons.
Since $H_C$ is invariant under the reflection $m \leftrightarrow K - m $, the non-degenerate eigenstates
have definite parity. This is the case in the two parton sector. Moreover, the $\mathbb{Z}_2$ symmetry
of planar states together with the "flavour" independence allows to identify the even and odd solutions
as identical boson-boson and identical fermion-fermion bound states. In the $K=\infty$ case, we have generated corresponding
eigenstates by employing bases with the required symmetry.

As usual, the lowest state is symmetric. It's wave function is constant in the parton momentum, $k$, with
exactly zero eigenvalue for any cutoff $K$
\footnote{Many other massless states have been found in the literature \cite{Harada:2004ck}
for any $K$ . These are due to a finite-$K$ artifact by which states with a large number of partons,
$k \sim K$, are annihilated by the SUSY charges (and therefore by the Hamiltonian), since all transitions
are blocked by the finite momentum resolution. Obviously only states with $k \ll K$ can be reliably
computed in our approach.}. This state is not displayed in the right panel of Figure \ref{ab-fg}.

Let us quote, for completeness, the continuum momentum limit of the
eigen-equations of $H_C$ which was used to obtain the $1/K=0$ points
in Fig. \ref{ab-fg}. They can be readily derived from the
discretized matrix elements (\ref{p2H}-\ref{p4H}) or, equivalently,
by applying our $H_C$ (\ref{Hquadratic},\ref{Hquartic}) to the
n-parton state and using rules of the planar calculus \cite{Veneziano:2005qs}. For
two partons one obtains:
\begin{equation}
 E_C \varphi(p)=  \frac{\lambda}{2\pi}\left[ \int_{\epsilon}^{p} dq
\frac{\varphi(p)-\varphi(p-q)}{q^2} + \int_{\epsilon}^{K-p} dq
\frac{\varphi(p)-\varphi(p+q)}{q^2} \right] \label{2partonH}\, ,
\end{equation}
which indeed is equivalent to 't Hooft equation:
\begin{equation}
 E_C \varphi(x) =  \frac{\lambda}{2\pi} PV \int_0^1 dy
\frac{\varphi(x)-\varphi(y)}{(x-y)^2} \label{tHeq}\,.
\end{equation}
It is perhaps worth observing that  in the former variables,
i.e. in terms of the momentum transfer  $q$, the linear {\em and}
logarithmic divergencies clearly cancel leaving behind the finite
right hand side of (\ref{2partonH}). This provides the justification
for the  principal value prescription commonly used in (\ref{tHeq}).

The multiparton generalizations follow from (\ref{Hquadratic},\ref{Hquartic}) as well

\begin{align}
&\notag E \varphi(p_1,  \dots p_n) =\\
&\label{npartons} \frac{\lambda}{2 \pi} \sum_{i=1}^n \int_{-p_{i+1}
+\epsilon}^{p_i - \epsilon} dq  \frac{ \varphi(p_1, \dots
p_i,p_{i+1}, \dots p_n) -  \varphi(p_1,  \dots p_i-q,p_{i+1}+q,
\dots p_n)}{q^2}\, ,
\end{align}
and, again, can be written in many equivalent forms \cite{Bhanot:1993xp}.
\subsection{Spatial structure of multiparton states}
\begin{figure}[h!]
\centering
\includegraphics[scale=.9]{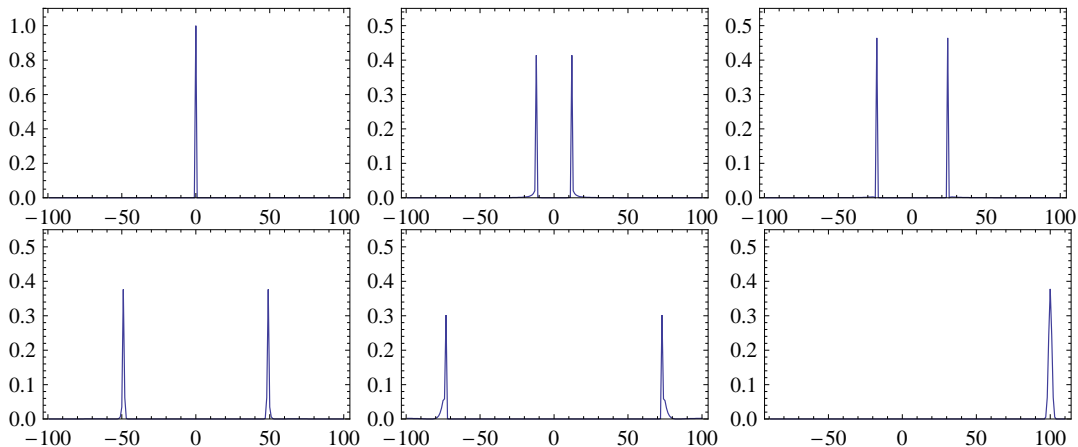}
\caption{Space profiles of various eigenstates (\ref{ft2p}) with two partons, K=201 $(r=1,26,50,100,150,200)$.}
\label{p2all}
\end{figure}
It turns out that the eigenstates of $H_C$ have a very simple and natural structure which shows up most
beautifully in position space. For $p=2$ it is summarized in Fig. \ref{p2all}
\noindent where a sample of two parton eigenstates spanning the whole interval of eigenenergies is displayed.
What is actually shown is the  modulus squared of the discretized version of the Fourier transform
\eqn
\psi_r(\Delta_{12}) = \int_0^P e^{-i \Delta_{12} p_1} \psi_r(p_1,P-p_1) d p_1,  \label{ft2p}
\eqnx
as a function of the relative, light cone, distance $\Delta_{12}=x_1-x_2$, with $\psi_r$ being the r-th
eigenstate of (\ref{p2H}). Upon the discretization
all momenta and coordinates become dimensionless integers: $P \rightarrow K$, $p_1\rightarrow m$,
$\Delta_{12}\rightarrow d_{12}$.

The interpretation of the above Figure is clear. The eigenstates are very well localized in relative distance,
and there is a very strong (linear in fact) correlation between $d_{12}$ and $r$.
In the lowest state two partons sit on
top of each other and their energy is exactly zero. Excited states correspond to partons separated by a
finite distance which gradually increases with the energy. Finally, in the highest state
partons are maximally separated, i.e. are located at the antipodes of the circle
(the Figure refers to  $K=201$). The relation between $d_{12}$ and eigenenergy turns out to be linear,
as expected in the one dimensional world (c.f. Sect.6.4). This also explains the linearity of the $p=2$ plot in
Fig. \ref{p234partonEnergies}.

\begin{figure}[h!]
\centering
\includegraphics[scale=.9]{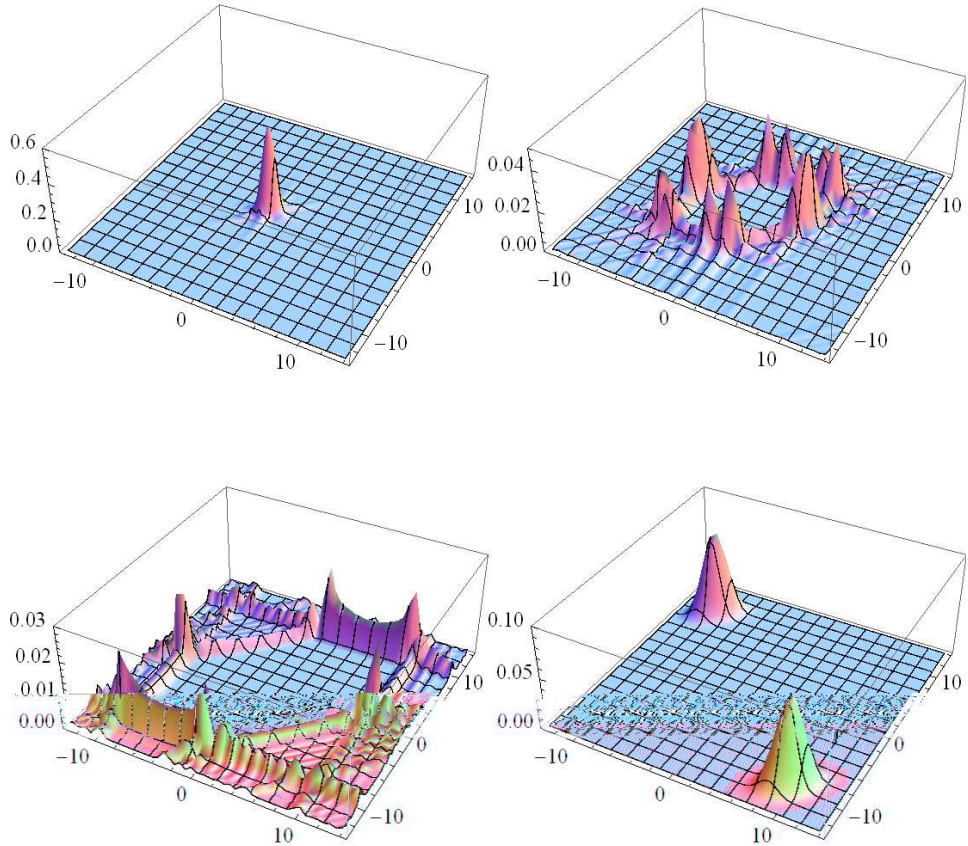}
%\begin{center}
%\epsfig{width=14cm,file=p3all.eps}
%\end{center}
 \caption{As in Fig.\ref{p2all}, but for three partons and in terms of two relative distances $d_{13}$ and $d_{23}$,
  $r=1,80,120,406$, $K=30$.}
  \label{p3all}
\end{figure}

Interestingly, this picture generalizes naturally to many parton sectors. Figure \ref{p3all}
displays a sweep through three parton eigenstates obtained by diagonalization of (\ref{p3H}).
Coordinate space wave functions depend now on two independent relative distances which we choose as
$\Delta_{13}=x_1-x_3$ and $\Delta_{23}=x_2-x_3$.
\eqn
\psi_r(\Delta_{13},\Delta_{23}) = \int_{p_1,p_2, p_3 > 0}^P e^{i \Delta_{13} p_1} e^{i \Delta_{23} p_2}
\psi_r(p_1,p_2,P-p_1-p_2) d p_1 dp_2\,.
\eqnx
Again, after the discretization the relative distances become integer: $ 1 \le d_{13}, d_{23} \le K-2 $,
with $K-2$ being the period of the discrete Fourier transforms.

Similarly to the $p=2$ case the wave functions are composed of a series of very narrow (in lattice units)
structures which give a sharp localization in relative distances. Again the energy of the lowest state
is exactly zero with all three partons  located at the same point (upper left panel). Going to higher and higher
states partons are moving apart migrating into the whole circle and finally, in the highest state,
three partons are sitting at the maximal and equal distances forming the familiar "mercedes" star (lower right).

Obviously the structure of three parton states is much richer
than that in the $p=2$ sector. Nevertheless a number of regularities
can be found which provide a compelling overall picture. They are
better seen in the contour plots which we will now discuss.

\begin{figure}[h]
\begin{center}
\epsfig{width=13cm,file=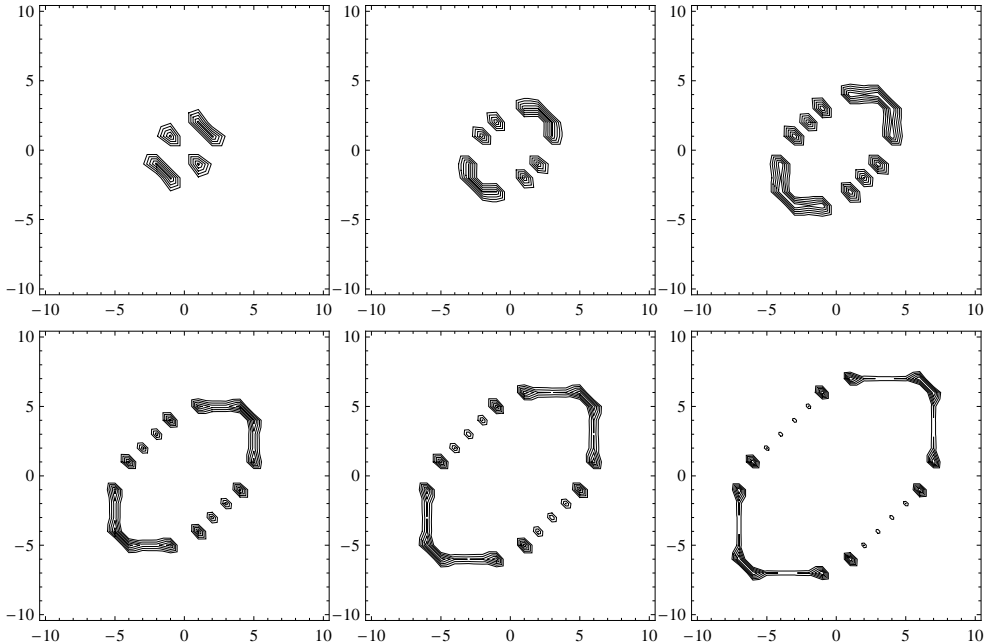}
\end{center}
\vskip-4mm \caption{A clean series (A) of three parton eigenstates,
$ K=100,\;\;\; 4 \le l \le 14 $.}
  \label{SeriesAframes}
\end{figure}

Apparently the whole spectrum consists of a series of states with
similar properties. Beginning of one such series is shown in Fig.\ref{SeriesAframes}. Again we see that the higher the energy, the larger are
the inter-parton distances. The new feature is that now there are
many peaks (i.e. parton configurations) in a single state.
Part of it comes from the $\mathbb{Z}_3$ symmetry of the displayed
densities.  However the rest provides a beautiful confirmation of the
linearity of the Coulomb potential and/or the underlying string
picture.  Namely, all parton configurations, composing a particular
state in one series, appear to have the same value of the "combined
string length" $l=|d_{12}|+|d_{23}|+|d_{31}| $. And vice versa:
configurations with different $l$ belong to states with different
energy. As an illustration consider the fifth state displayed in
Fig. \ref{SeriesAframes}. It has $l=12$, this can be achieved, e.g.
with partons 1 and 2 separated by 6 units and with parton no. 3
somewhere between the two. This is represented by 5 configurations
(peaks) extending from $(d_{13}, d_{23})=(1,-5)$ to $(5,-1)$ as no. 3
moves from no. 1 to no. 2. Similarly one can decode all structures
appearing in that Figure. One more example: a horizontal ridge
extending between $(-5,-6)$ and $(-1,-6)$, in that panel, describes
partons 2 and 3 separated by 6 units and parton 1 located in five
positions between the two.

\begin{figure}[h]
\begin{center}
\epsfig{width=13cm,file=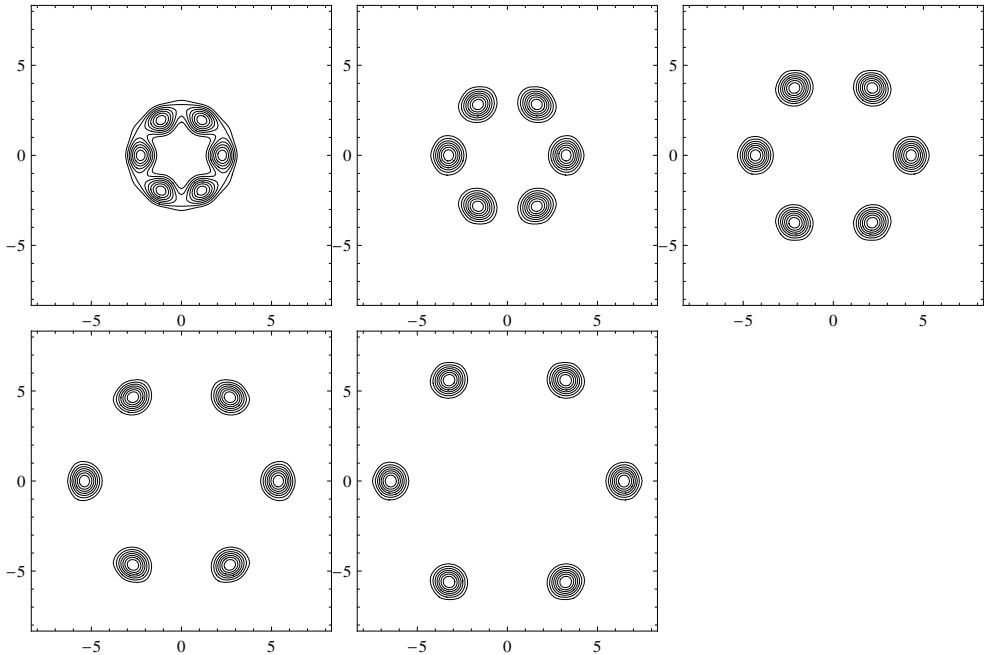}
\end{center}
\vskip-4mm \caption{Different series (D) and on the ``Dalitz plot".}
  \label{DalitzD}
\end{figure}

The $\mathbb{Z}_3$ symmetry of planar states is not evident in these figures, because of the asymmetric variables
used. However it is there and corresponds to
$(x,y)=(d_{13}, d_{23}) \rightarrow ( y-x, -x) \rightarrow (-y,x-y)$. The $\mathbb{Z}_3$ symmetric
representation is shown in Fig.\ref{DalitzD} where another series is displayed on the massless ``Dalitz plot"
ensuring the constraint $d_{12}+d_{23}+d_{31}=0$.

Finally, let us comment on the existence of different series in the many body spectrum
which we find quite intriguing. As already said, we see "experimentally" that our three parton states
group naturally into series. States in one series differ only by the increase of the relative distance between
partons, as in Fig.\ref{SeriesAframes}. However states from various series exhibit other differences.
An example of another series was just given in Fig.\ref{DalitzD}. Here again, increasing the
inter parton distances increases the eigenenergy, however the pattern of the configuration remains the same.
On the other hand, apart from the change of a display, there is a clear difference between patterns shown in Figures \ref{SeriesAframes}
and \ref{DalitzD}. In the first series (A) the partons never coincide, while in the other one (D)
in every configuration one relative distance vanishes. However, we have also found other series where the
differences are not so clear.
In general, states from different series can have the same combined string length $l$
and yet they have different energies. This suggests the existence of other conserved quantities,
hence quantum numbers, which also control multiparton spectra.

\begin{figure}[tbp]
\begin{center}
\epsfig{width=12cm,file=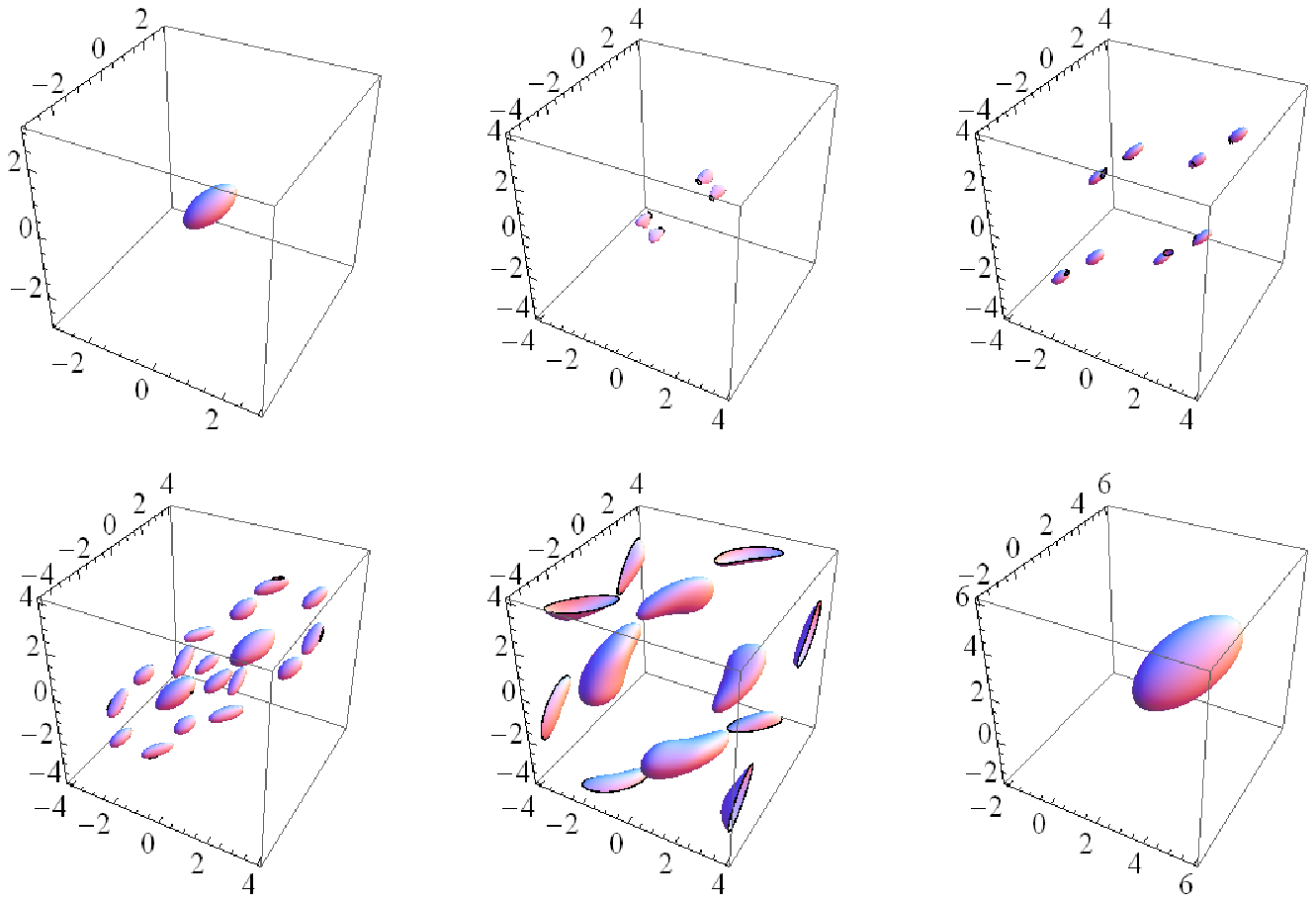}
\end{center}
\vskip-4mm \caption{Structure of eigenstates with four partons.
Contour plots in three relative distances ($d_{14},d_{24},d_{34}$)
for states no. 1,9,35,60,100,165 spanning the whole range  of
states for $K=12$, $r_{max}=165$.}
 \label{p4all}
\end{figure}

Solutions in the four parton sector show qualitatively the same
phenomena. In Fig \ref{p4all} we display contour plots, in three
relative distances $d_{14},d_{24}$ and $d_{34}$, of various
eigenstates of (\ref{p4H}). Again, in the lowest state all
relative distances vanish;  then, in higher states, partons gradually
separate and in the highest state they group in two closely bound pairs
sitting at the antipodal locations on the circle. This is different
than the naive extrapolation from the three parton case and illustrates
how rich is the system with possibly more surprises at higher multiplicities.
Similarly to $p=2,3$, states
are sharply localized in the relative distances and separations
between various partons can be determined.

Obviously, multidimensional representations like Fig.\ref{p4all} become unpractical
for higher multiplicities. An alternative way to proceed is to study inclusive densities and correlations.
In our case, e.g. in a given $p$ parton sector, an inclusive single parton density can be defined as
\eqn
D_r(\Delta)= \int d^{p-1} \vec{\Delta_{p}} \sum_{i=1}^{p-1} \delta(\Delta - \Delta_{in}) |\psi_r(\vec{\Delta}_{p})|^2
\eqnx
and gives the number of partons at a distance $\Delta$ from i.e. the last one. It can be easily
calculated from our exclusive wave functions or, yet simpler, directly from the Fourier components.
The latter representation reads, e.g. in the four parton case,
\eqn
D_r(\Delta)=\int_{p_2,p_3,P-p_2-p_3 > 0}^P dp_2 dp_3 |\psi_r(\Delta,p_2,p_3)|^2 + cycl.
\eqnx
with $\psi_r(\Delta,p_2,p_3)$ standing for the partial Fourier transform - only in the first variable.

\begin{figure}[tbp]
\begin{center}
\epsfig{width=12cm,file=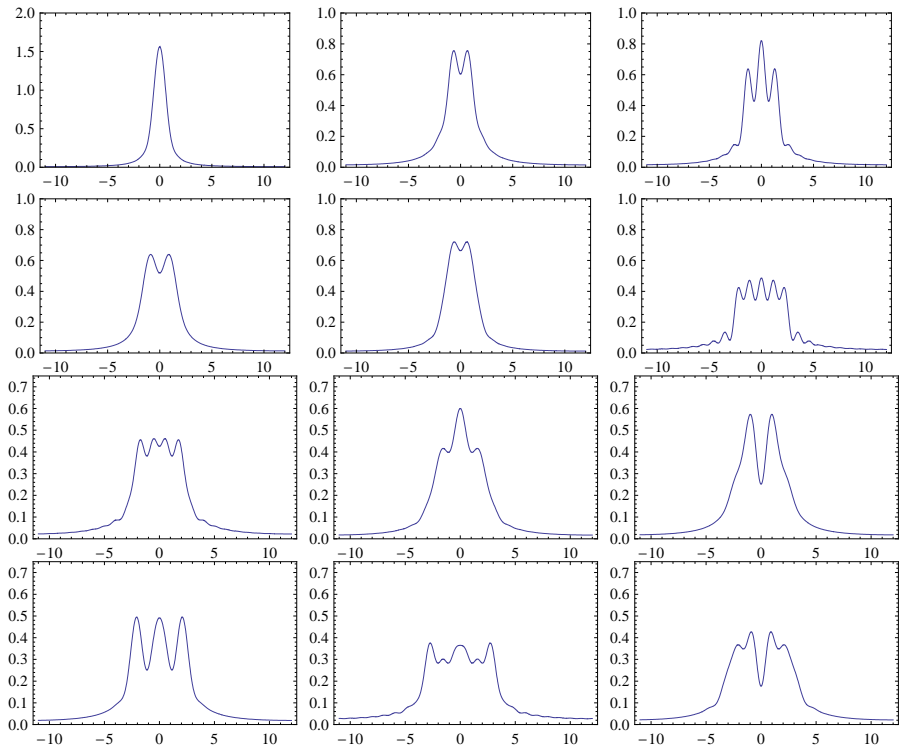}
\end{center}
\vskip-4mm \caption{ Inclusive single parton density for four partons and for lower
states $r=1,4,5,6,9,12,13,14,15,20,26,29$, $K=27,\;\; r_{max}=2600$.}
 \label{p4lower}
\end{figure}

In Fig. \ref{p4lower} above density is shown in the four parton
sector. It confirms what we have already learned from the exclusive data. However
the structure of four parton states can now be seen in more detail. Rather than sweeping through
the whole spectrum, we have concentrated on lower states. Fig.\ref{p4lower} clearly shows
the growth of the distance between the two outermost partons and how the intermediate positions
between the two are populated. Of course the complete information
could be recovered only upon examining simultaneously higher
inclusive densities.

\subsection{Some analytic considerations}

Let us now discuss some analytic  aspects of the solutions. To this  date such solutions are not available
in spite of many attempts \cite{Fateev:2009jf}. Generalization for many bodies only increases the challenge.
We believe however that the numerical studies presented here can
also contribute to the analytic understanding of these systems. Consider for example the massless, $E_C=0$,
bound state. It has been found numerically, and it is obvious from (\ref{npartons})
that such a solution exists for all multiplicities and has a constant wave function in the momentum
representation. It is then a simple matter to construct its (LC) configuration space counterpart.
For two partons we have
\begin{align}
 \psi(x_1,x_2) &\notag= c \int_0^P dp_1 dp_2 e^{-ix_1 p_1-ix_2 p_2} \delta(p_1+p_2-P) \\
&=  e^{-i P(x_1+x_2)/2} \frac{\sin(P\Delta_{12})}{\sqrt{P}\Delta_{12}},
\end{align}
with the (not normalized) probability distribution depending only on $\Delta_{12}$
\begin{equation}
 |\psi|^2= \frac{\sin^2(P\Delta_{12})}{P\Delta_{12}^2}\,.
\end{equation}
For three partons one obtains analogously
\begin{align}
 \psi(x_1,x_2,x_3) &\notag= c \int_0^P dp_1 dp_2 dp_3\, \,\delta(p_1+p_2+p_3-P)
 e^{-ip_1 x_1-i p_2 x_2-i p_3 x_3}=\\
 &=i c \frac{ e^{-i x_3 P}}{2 \Delta_1 \Delta_2 (\Delta_1-\Delta_2)} \left(
\Delta_1 e^{-i \Delta_2 P} \sin(\Delta_2 P) -\Delta_2 e^{-iP\Delta_1} \sin(\Delta_1 P)
\right)\, ,
\end{align}
where we used $\Delta_1=\Delta_{13}/2,\Delta_2=\Delta_{23}/2$ as relative
distances. Upon normalization $c= P/\sqrt{2}$, we obtain for the density
\begin{align}
 |\psi|^2=\frac{P^2}{32 \Delta_1^2 \Delta_2^2 (\Delta_1-\Delta_2)^2}&\notag\left[
 (\Delta_1 \sin(2 \Delta_2P)-\Delta_2 \sin(2\Delta_1 P))^2 +\right.\\
&\label{3partonsWV}\phantom{a} \left. + 4 (\Delta_1 \sin^2(\Delta_2
P)-\Delta_2 \sin^2(\Delta_1 P))^2 \right]\,,
\end{align}
which is completely symmetric in exchanging $\Delta_1$ with
$\Delta_2$ and regular for $\Delta_1,\Delta_2 \rightarrow 0$. As for
the two parton case we see that in the limit $P\rightarrow \infty$
we find a $\delta$-function  centered at zero, both for $\Delta_1$
and $\Delta_2$, confirming that actually the zero energy state
corresponds to three partons sitting all together at the same
position.

Generalizing this computation to higher sectors is trivial, hence the massless
eigenstates can be constructed analytically in all multiplicity sectors.

\subsection{String picture}

The  findings reported so far  support the widely accepted string picture of (1+1)- dimensional
planar gauge theories. The relation is quite natural, even hardly surprising, in the case
of two partons. It is also generally expected with more partons, however our results provide a
direct illustration of  how this is happening.

\begin{figure}[h]
\begin{center}
\epsfig{width=8cm,file=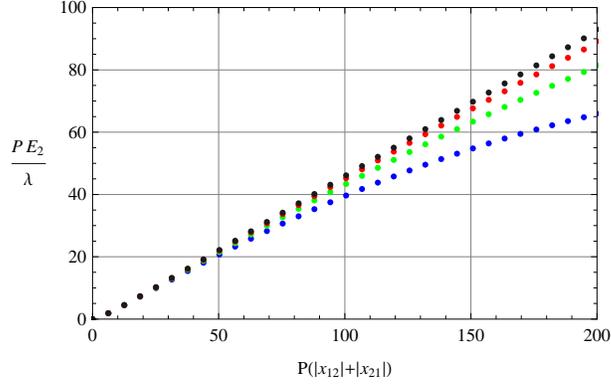}
\end{center}
 \caption{Two-parton energies, as a function of (doubled) parton separation,  for increasing $K$.
  Different colors (bottom to top) correspond to $K=50,100,200,400$.}
  \label{E2(l)}
\end{figure}

We have seen that the two parton eigenstates are very well localized in  (LC) configuration space.
Consequently, the length of the effective string between two partons can be readily extracted from
our eigenstates (c.f. Fig.\ref{p2all}). Figure \ref{E2(l)} shows the dependence of the eigenenergies
on that length. It approaches a nice linear form at large cutoff, $K$, with a well defined,
finite, string tension.

There is an interesting correlation between the parity under the reflection $ k \leftrightarrow P - k $,
and the inter parton distances. Namely, in even (odd) states partons are separated by integer (half-integer)
distances. This explains why the eigenenergies of the symmetric and antisymmetric states are half a way
between each other in Fig. \ref{ab-fg}

\begin{figure}[h]
%\begin{center}
\epsfig{width=8cm,file=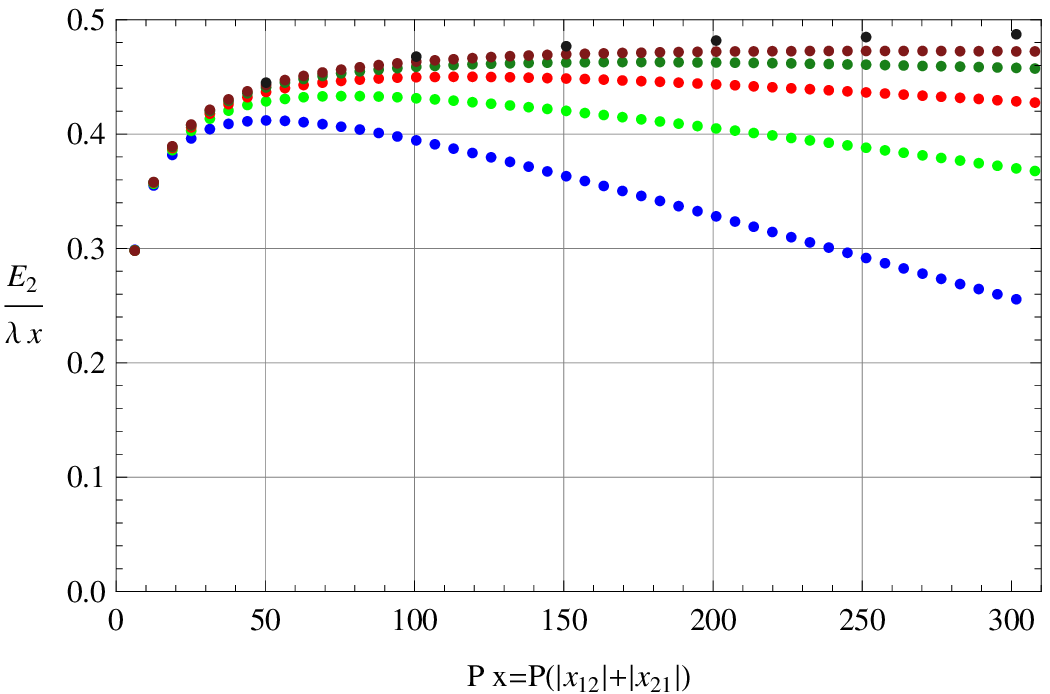}
\epsfig{width=8cm,file=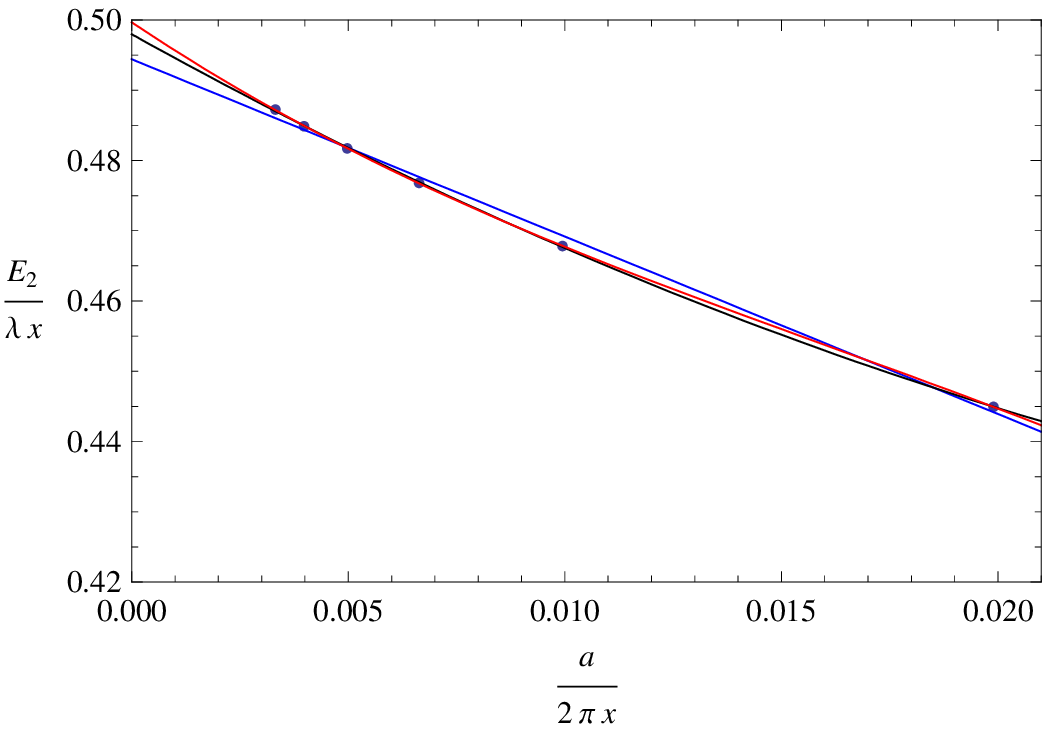}
%\end{center}
 \caption{Left: as in Fig.\ref{E2(l)} but for the string tension
 together with the extrapolations to $K=\infty$ (black). Right: polynomial extrapolations (see the text) to $a=0$.}
  \label{tension1}
\end{figure}

Before moving to higher multiplicities, let us compare  our
numerical results with the theoretical prediction for the effective string
tension in the two-parton case, $\sigma=\lambda/2$ \cite{Kazakov:1980zi, Kazakov:1980zj}.  Fig.\ref{tension1} shows the ratio of two
parton eigenenergies to the combined string length
$x=|x_{12}|+|x_{21}|$, in units of $\lambda$, as a function of the
dimensionless lattice distance $P x = 2\pi x/a = 2\pi l$. Different
colors show results for different cutoffs, $K=50, 100, 200, 400, 800$.


 As
expected, there is a significant cutoff dependence for larger parton
separations. The K dependence is however rather weak and can be
easily taken care of. Black points show results of the polynomial
(in $1/K$) extrapolation to $K=\infty$, at few values of $l$. In
lattice terminology they correspond to the string tension determined
from finite lattice distances hence are still biased by finite $a$
effects. To get rid of the latter one has to perform the continuum
(i.e. $a\rightarrow 0$) limit. This is summarized in
Fig.\ref{tension1} where the $a/x$ dependence of above K
extrapolations is displayed. Again the $a/x$ dependence is mild and
polynomial fits provide quite stable extrapolations to $a=0$. Those
are in very satisfactory agreement with the theory cf. Table I.
\begin{table}[h!]
\begin{center}
\begin{tabular}{cccc}
 \hline\hline
  $M$            &  $   1   $ & $  2  $ &  $ 3   $  \\
   %\hline
  $W_{M}(0)$   &  $  0.4944  $ & $ 0.4980 $ &  $ 0.4997 $ \\
   \hline\hline
\end{tabular}
\end{center}
\caption{First three polynomial extrapolations of the $a$ dependence
of the string tension from Fig.\ref{tension1}.} \label{conti}
\end{table}

\begin{figure}[h]
%\begin{center}
\epsfig{width=8cm,file=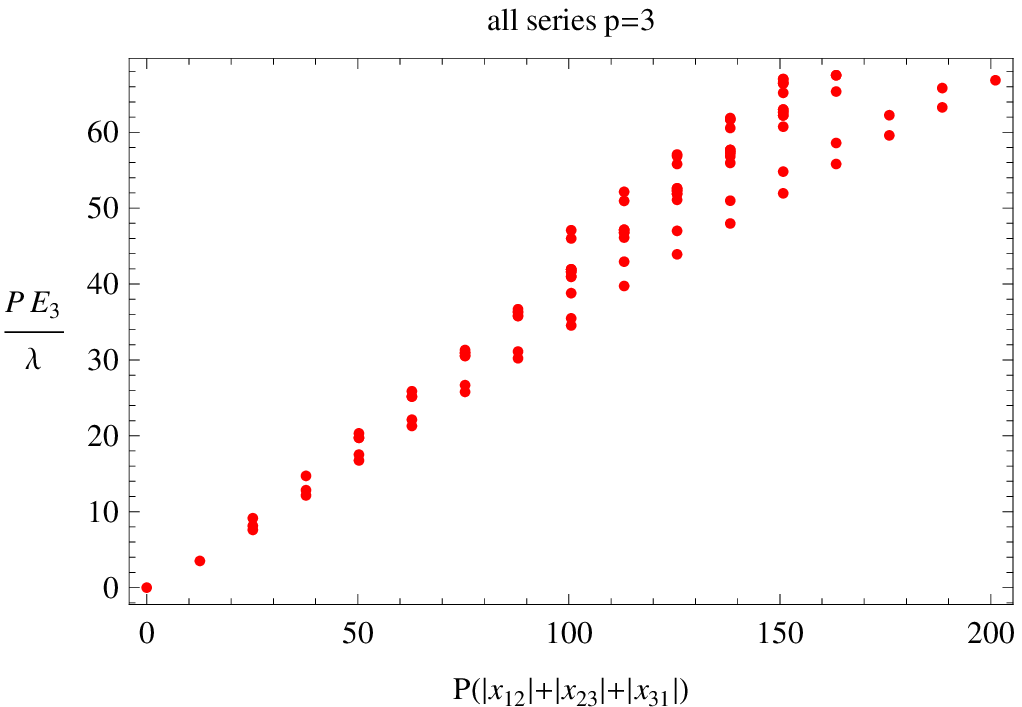}
\epsfig{width=8cm,file=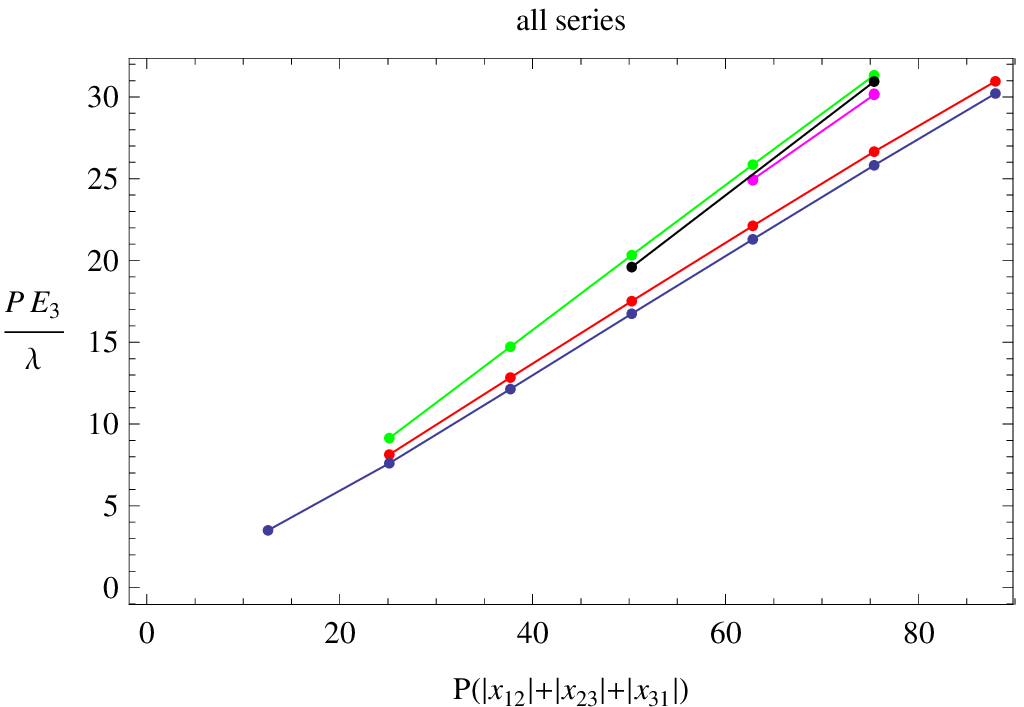}
%\end{center}
\vskip-4mm \caption{Eigenenergies of three parton states vs. the combined string length $l$.}
  \label{p3E(l)}
\end{figure}

With three partons, situation is yet more interesting. Assigning
automatically the combined string length $l$ to each state on the
basis of integer coordinates of sharp peaks in the density, results
in Fig. \ref{p3E(l)} (left, $K=100$). This only confirms the existence of some
ambiguities. We have already found, however, that there are series
of states composed of similar patterns of configurations. Could they
account for what is seen in Fig. \ref{p3E(l)}? In fact yes, in the right panel of Fig.
\ref{p3E(l)} we show similar plot but
constructed for five series identified by inspection of lowest 20
eigenstates. Two of them were described in the previous subsection.
Indeed, for each series we observe a clean linear growth of the
energy with $l$. The slope of that dependence is consistent (albeit somewhat smaller) with the
one seen for two partons.

\begin{figure}[h]
\begin{center}
\epsfig{width=8cm,file=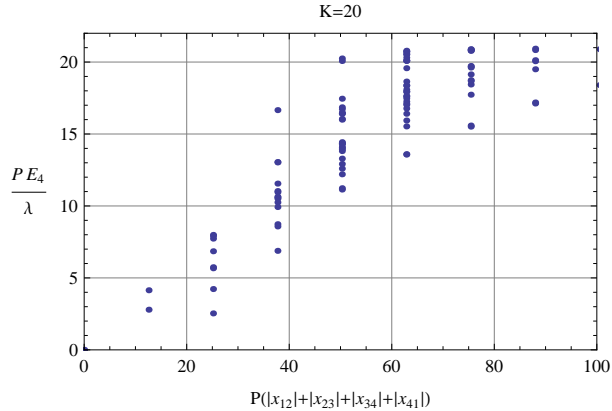}
\end{center}
\vskip-4mm \caption{As in Fig.\ref{p3E(l)} but for four
partons.}
  \label{E4(l)K20}
\end{figure}

Similar behaviour is seen in the four parton sector, Fig. \ref{E4(l)K20},
obviously the situation is more complex
with more families and larger spread of energies at fixed "l". No attempt was made to identify
separate series yet and to extract the string tension. The latter requires the former and
extrapolations in $K$ and $a$, as was done for $p=2$.

Clearly more detailed studies are needed to unravel a complete structure of higher parton states.
In particular comparison with the direct solutions of 't Hooft equations in higher-$p$ sectors would
be very useful.
\subsection{Pre-SUSY}
As already mentioned our drastic approximation breaks half of the  original supersymmetry leaving behind just the ${ \cal N}(1,1)$ subgroup generated by   $Q_2$,   $Q_2^{\dagger}$.
In spite of this it is amusing to compare the spectra of sectors that should be connected by the action of all the  SUSY charges.
The supersymmetry generated by $Q_2$,    $\bar{Q}_2$ connects states with the same number of partons and is self-evident in our graphs. On top, there are further degeneracies between states with the same $p$ due to our Coulomb approximation. It is clear, instead, that there is no exact degeneracy for states of different $p$ even if some of them should be connected through the action of the  $Q_1$,    $\bar{Q}_1$ generators.

\begin{figure}[h]
\begin{center}
\epsfig{width=12cm,file=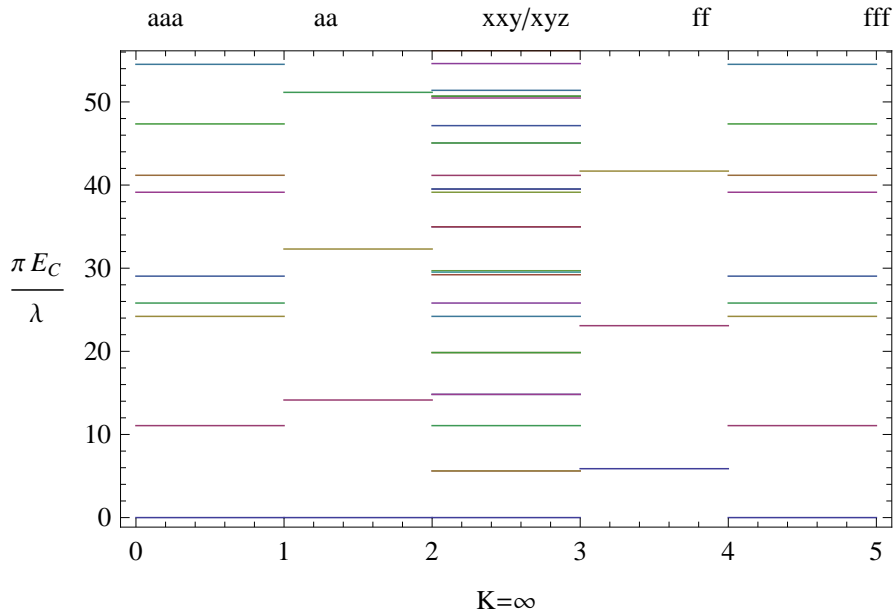}
\end{center}
 \caption{The lowest levels with 2 and 3 partons. Here $a$ represents either $a$ or $b$ and $f$ either
 $f$ or $g$. For $p=2$ the $aa$ and $ff$ spectra, for identical partons, are complementary and their sum
 gives all the unequal-parton spectra. For $p=3$ the $xxx$ (i.e.  $aaa$ or $fff$) spectra are all identical
 but contain far less states than the $xxy$ or $xyz$ spectra. The $xyz$ spectra are equal to the $xxy$
 spectra modulo a doubling of the states since $xyz$ and $xzy$ are two distinct, but degenerate, states.}
  \label{preSUSY}
\end{figure}

Let us consider, as an example, the Konishi (anomaly) supermultiplet keeping Fig. \ref{preSUSY}  in mind.
As discussed in section 2.1 the full chiral supemultiplet in this case contains $p=2$ states of
non-identical partons and $p=3$ states in which there are at least two different species.
The corresponding energy levels match the union of the $aa$ and $ff$ spectra in the figure
and the central ($xxy/xyz$) levels. We see that there is (decreasingly) good matching for the first
three excited levels while  for the fourth the matching is already quite poor
(for higher levels the good matching looks a bit accidental given the high density of $p=3$ levels).

Consider now a different supermultiplet, one that contains the state $|aa\rangle$.
Its partners should be found in the $af$ sector and in the $aaf$ and $aff$ sectors
and all works as in the previous case.

Consider finally the supermultiplet containing the state $|ff\rangle$.
Apparently we find a problem since the lowest excited $ff$ state has no nearly-degenerate partner
in the $fff$ sector. However this is as it should be: the $|ff\rangle$ wavefunction must be odd
under the interchange of the two momenta. When one applies $ Q_1$ to it one finds
that this antisymmetry clashes with the symmetry needed in $|fff\rangle$.
Instead, the $|ff\rangle$ state should form a multiplet with $|af\rangle$, $|afg\rangle$ and $|abf\rangle$.
The matching is now excellent.

In conclusion, although a priori our drastic truncation of the Hamiltonian could have left no sign at all of
the supersymmetry generated by $Q_1$ and    $\bar{Q}_1$, we have found that some trace of the full
${\cal N}=(2,2)$ supersymmetry appears to have survived in the spectra of $H_C$.
This makes us confident that our truncated Hamiltonian  may represent a fairly good
approximation to the exact one.

\section{Conclusions and outlook}

We have considered the dimensional reduction of $D=4,\, {\cal N} =1$ SYM  theory to two dimensions in the large-$N$ (planar) limit using the LC  gauge and LC quantization.
This allows us to explicitly eliminate all non-physical degrees of freedom at the price of having a non-local LC Hamiltonian. Nonetheless, such an Hamiltonian exhibits many desirable features: it is invariant under  ${\cal N} =(2,2)$ supersymmetry transformations provided these are suitably defined  in order not to destroy the LC-gauge choice; it is manifestly normal-ordered and positive-semidefinite  so that it possesses an exact zero-energy ground state and a spectrum of non-negative energy excitations.

In order to solve for the eigenvalues and eigenstate of the Hamiltonian we have
compactified LC-space to a circle of radius $R$ allowing to work within a finite Hilbert
space as long as we keep the conserved LC momentum of our states finite, the idea being, of course,
to eventually send $R$ to infinity and check that physical quantities approach a finite smooth limit.

Infrared (IR) divergences appear to make this task somewhat technically complicated
(although in principle possible) and therefore, in this first paper, we have truncated
the Hamiltonian to what looks superficially as its most IR divergent part.
Indeed, for the colour-singlet, single-trace states that survive in the large-$N$ limit,
these linear IR divergences are neatly cancelled and get replaced by an effective Coulomb
interaction which, in $D=2$ gives a confining linear potential.
We have then studied many properties of the eigenvalues and eigenfunctions of this Coulomb Hamiltonian,
$H_C$, and confirmed that, at least in this approximation, the large $R$
limit is smooth (although larger and larger $R$ are needed before heavier and heavier states stabilize).

The resulting model breaks supersymmetry down to an ${\cal N} =(1,1)$ subgroup and  looks like
a supersymmetric generalization of 't Hooft's original model \cite{'tHooft:1974hx}  where
states with an arbitrary number of partons are present even at leading order in $1/N$.

When seen in position-space a nice string picture emerges in which the mass of each state
is proportional to the sum of the (center-of-mass distances) of each pair of neighbouring partons.
 These distances are quite sharply quantized leading to a discrete
spectrum with approximately linear ``Regge" trajectories. The
numerical value of the proportionality constant (the string tension)
agrees very well with theoretical expectations. For states
with more than two partons the situation is obviously richer. Yet we were able to identify
clean series of three-parton states whose eigenenergies are indeed proportional to the combined
length of strings stretched between neighbouring partons. Moreover, the detailed patterns
of parton configurations contributing to these states confirm unambiguously
the linear form of the two-body interactions. String tension seen
in these sectors is compatible with the one extracted from the two parton sector.

In our Coulomb approximation the first excitation over the Fock vacuum is also massless with
all the partons sitting at the same point but this is most likely an artifact of our Coulomb approximation
that allows all partons to sit at the same point without paying any kinetic-energy price. Since even those massless states are nicely paired in supermultiplets,
we expect them to be lifted to some finite energy.  Indeed, if we  add some finite terms present in the
full Hamiltonian (\ref{HamiltIO}),  we see that these state acquire a non-zero energy.
At least in our approximation we find no evidence (apart from the just mentioned states)
for the absence of a mass gap reported in some previous studies \cite{Harada:2004ck}.
We also see clearly how the existence of many other massless states, for each value of $R$,
is nothing but a consequence of the breakdown of the method when the number of partons approaches
its maximal value compatible with momentum conservation.

Finally, we found that, even in our drastic approximation the full ${\cal N} =(2,2)$
supersymmetry of the original model shows up as an approximate supermultiplet structure
at least for the lightest states.
Our belief is that the ``worst'' divergent part in (\ref{HamiltIO}) gives us
the bulk structure of the energy states,
namely the nice discrete linear spectrum. We expect  the logarithmic IR singularities
to lead to a dressing of our states \`a la Block-Nordsieck and to modify  the energies by lifting,
in particular, the zero-energy states.
We plan to report on progress in this direction in a forthcoming paper.

\section*{Acknowledgements}
Two of us (DD and JW) would like to thank the organizers and the audience of the workshop on Large-$N$ Gauge Theories at  the Maryland Center for Fundamental Physics, University of Maryland, May 2010, for hospitality and discussions.
DD would  like to thank the  high energy physics group of DAMTP, Cambridge UK, where he was visiting as Ph.D. student during the write up of this paper.
GV would like to acknowledge  illuminating discussions with  Adi Armoni, Giancarlo Rossi and Adam Schwimmer.
This work is supported in part by the International PhD Projects Programme of the Foundation for Polish Science within 
the European Regional Development Fund of the European Union, agreement no. MPD/2009/6.

\section*{Note Added}

After having circulated a preliminary version of this paper we became aware of previous work \cite{Gross:1995bp}, \cite{Armoni:1997ki}, \cite{Gross:1997mx}, 
\cite{Armoni:1998kv}, claiming that two-dimensional theories of the kind we considered (i.e. with massless fermions in the adjoint representation) should exhibit a Schwinger-like phenomenon even in the planar large-$N$ limit. This would result in the screening of the linear potential and in the spectrum becoming continuous above a certain energy scale. If this were true, it would mean that the terms we neglected should have a dramatic effect, at least for the high-energy part of the spectrum. This point, as well as the dependence of the spectrum itself on the scalar VEV, clearly deserve further investigations. We are grateful to A. Armoni for bringing the abovementioned papers to our attention.

\bibliography{DVW2arch}{}
\bibliographystyle{ieeetr}

\end{document}